\documentstyle[pre,preprint,aps]{revtex}   

\tighten
\begin{document}
\input epsf  
\preprint{Draft \today}
\draft

%
%
%
%
%
%

\title{Nonlinear dynamics of a solid-state laser with injection}
\author{M. K. Stephen Yeung\thanks{Corresponding author;
				   email: yeung@tam.cornell.edu}
and
Steven H. Strogatz\thanks{email: strogatz@cornell.edu} }
\address{Department of Theoretical and Applied Mechanics, \\
	 Kimball Hall, Cornell University, Ithaca, NY 14853-1502}
\date{\today}
\maketitle

\begin{abstract}

We analyze the dynamics of a solid-state laser driven by an injected
sinusoidal field.  For this type of laser, the cavity round-trip time 
is much
shorter than its fluorescence time, yielding a dimensionless ratio of
time scales $\sigma \ll 1$.  Analytical criteria are derived for the
existence, stability, and bifurcations of phase-locked states. We
find three distinct unlocking mechanisms.  First, if the dimensionless
detuning $\Delta$ and injection strength $k$ are small in the sense
that $k = O(\Delta) \ll \sigma^{1/2}$, unlocking occurs by a
saddle-node infinite-period bifurcation.  This is the
classic unlocking mechanism governed by the Adler equation:
after unlocking occurs, the phases of the drive and the laser drift
apart monotonically.  The second mechanism occurs if the detuning and
the drive strength are large: $k =O(\Delta) \gg \sigma^{1/2}$.
In this regime, unlocking is caused instead by a supercritical Hopf
bifurcation, leading first to phase trapping and only then to phase 
drift as the drive is decreased.  The third and most
interesting mechanism occurs in the distinguished intermediate regime
$k, \Delta = O(\sigma^{1/2})$.  Here the system exhibits complicated,
but nonchaotic, behavior.  Furthermore, as the drive decreases below
the unlocking threshold, numerical simulations predict a novel
self-similar sequence of bifurcations whose details are not yet 
understood.
\end{abstract}


\narrowtext


\section{Introduction}

The Adler equation
\begin{equation}
   {d \Phi \over dt} = \Delta - k \sin\Phi
\label{PLL}
\end{equation}
provides the simplest model of phase locking between a nonlinear
oscillator and an external periodic drive.  Here $\Phi(t)$ is the
phase difference between the oscillator and the drive, $\Delta$
is the frequency detuning, and $k$ is the coupling strength.
This equation first arose in connection with the  phase locking of
microwave oscillators \cite{Adler46}, and has since found application
in many other settings, including the depinning  of
charge-density waves \cite{Gruner83}, the entrainment of biological
oscillators \cite{ER84,Strogatz87},
and the onset of resistance in superconducting Josephson junctions
\cite{Stewart68,Strogatz94}.

A system governed by the Adler equation can display only two types of
long-term behavior \cite{Strogatz94}.
If $|\Delta/k| \leq 1$, all solutions tend to a phase-locked state,
where the response oscillator maintains a constant phase difference
relative to the driver.  On the other
hand, if $|\Delta/k| > 1$, all solutions exhibit phase drift, where the
phase difference grows monotonically, with one oscillator periodically
overtaking the other.

The main limitation of the Adler equation is that it treats the
response oscillator as a system with only one degree of freedom,
namely its phase.  Possible variations in its amplitude (and any
other degrees of freedom) are ignored.  This approximation is
reasonable in the limit of weak driving; in
that case, the amplitude of the response oscillator typically
equilibrates much more rapidly than its phase, and can therefore be
treated as a constant in the subsequent analysis.  But if the driving
is not weak (in some appropriate dimensionless sense), the dynamics
can become complicated.  In this paper we revisit a classic problem
-- the mathematical analysis of
a solid-state laser with external injection
\cite{Siegman86,WV91,BFS73,Byer93,Barillet96} -- and explore it
in regimes where amplitude effects
become important and the Adler approximation breaks down.

Our work was inspired by recent theoretical and experimental studies of
amplitude effects in two mutually coupled solid-state Nd:YAG lasers
\cite{EKC95,TMRCLE97}.
In those studies, the lasers were equally
coupled and identical, except for a slight relative detuning of their
frequencies from some common cavity mode.  For coupling strengths well
above or below the locking threshold, the lasers were found to exhibit
the simple behavior expected from the Adler approximation.  However,
as the coupling approached the locking threshold from below, the
lasers showed a series of amplitude instabilities, culminating in a
period-doubling route to chaos.  These novel instabilities could
not be explained by the Adler approximation.  Instead the authors
proposed the following mechanism.  Below the locking threshold,
the lasers exhibit phase drift.  If the time required for one full
cycle of phase slip happens to be an integer multiple of the lasers'
relaxation period, the resulting subharmonic resonance might account
for the observed instabilities.  For the highly symmetrical case
where the two lasers are assumed to have identical intensities and
gains, this argument was proven to be correct by reducing the
governing equations to those for a single, periodically modulated
laser, where the subharmonic resonance mechanism was already known
to occur \cite{EBM87,SE94}.

We wondered whether similar amplitude instabilities and chaos would
occur in two coupled Nd:YAG lasers with {\it unidirectional} coupling
(or equivalently, in a single Nd:YAG laser with external injection).
On the one hand, the qualitative argument about subharmonic resonances 
should still work.  On the other hand, the equally-coupled case enjoys 
special symmetries that are not present in the unidirectional case.  
Given the crucial role of the symmetry in the earlier analysis of 
Erneux {\it et al.} \cite{EKC95,TMRCLE97}, it seemed possible that 
some new effects might occur if the symmetry were broken.

As we will show below, the system with one-way coupling can indeed
display some fascinating behavior near the unlocking threshold, but
it differs from that seen in the equally-coupled case.  In particular, 
we do not see a period-doubling route to chaos, nor any evidence of 
chaos at all.  Instead, in a certain distinguished regime of 
parameters, we find a self-similar cascade of periodic windows and 
bifurcations.  To the best of our knowledge, this bifurcation 
scenario is novel.

It will be interesting to see whether this cascade can be detected
experimentally for a laser in the appropriate parameter regime, as
specified by our theory.  It would also be gratifying to have a better
theoretical understanding of the cascade itself.

This paper is organized as follows.
The governing equations are given in Sec.\ \ref{section_Model}.
In Sec.\ \ref{section_Nondim_Sym}, we reduce the number of parameters
by nondimensionalizing the equations and exploiting certain
symmetries.  By choosing a frame that co-rotates with the phase of
the driver, we reduce the system  to three coupled autonomous ordinary
differential equations: one for $\Phi(t)$, the
phase difference between the laser and the drive, and one each for the
dimensionless gain and amplitude of the response laser.  Fixed points
of this reduced system correspond to injection-locked states of the
original system.  Sec.\ \ref{section_Special_Cases} dispenses with the
limiting cases of zero coupling or zero detuning where the dynamics
can be analyzed straightforwardly.

The analysis begins in earnest in
Sec.\ \ref{section_Locked_States}, where we derive criteria for the
existence and stability of injection-locked states, and compare our
criteria to those obtained in the usual Adler approximation.
In Secs.\ \ref{section_Weak_Coupling} and
\ref{section_Strong_Coupling}, we start to investigate what happens
when locking is lost.  We show perturbatively that for a broad range
of parameters, the phase difference $\Phi(t)$ oscillates periodically,
but the precise nature of those oscillations depends on the relative
sizes of the dimensionless coupling, detuning, and stiffness of the
system.

For a distinguished limit of parameters, described in
Sec.\ \ref{section_Beyond_Periodicity}, the reduced system has
complicated dynamics and undergoes the self-similar cascade of periodic
windows and bifurcations mentioned above.  In 
Sec.\ \ref{section_A_Singular_Limit}, we consider the system in the 
singular limit of zero stiffness.  Again, the cascade persists.  Based 
on the distinctive helical structure of the periodic orbits, we propose
a mathematical mechanism underlying the cascade.  It apparently stems 
from a codimension-two bifurcation in which a supercritical Hopf 
bifurcation combines with a saddle-node infinite-period global 
bifurcation.  We have no proof of this mechanism, but show that it
correctly predicts the scaling laws found numerically in the 
bifurcation diagram.  We conclude with a discussion of open questions.


\section{Formulation of the Model}
\label{section_Model}

For solid-state lasers, as well as other Class~B lasers with
negligible linewidth enhancement factors, such as $\mathrm{CO_2}$ and
ruby (NMR) lasers \cite{WV91},
the polarization relaxes rapidly compared to the electric field and the
gain, and can therefore be adiabatically eliminated.  Following
Ref.~\cite{FCRL93} with straightforward modifications, we can write the
following equations for two lasers coupled through transverse overlap
of their electric fields, assuming single-mode operation and
neglecting spatial variations within the lasers:
\begin{eqnarray}
  {d E_1 \over d T^*} &=& \tau_{\mathrm{c,1}}^{-1} [ (G_1 - \alpha_1)
			  E_1 + \mu K E_2'] + i \omega_1 E_1 +
			  \sqrt{\epsilon_1} \xi_1 (T^*),
		          \nonumber \\
  {d G_1\over d T^*} &=& \tau_{\mathrm{f,1}}^{-1} (p_1 - G_1 -
			 G_1 |E_1|^2), \nonumber \\
  {d E_2 \over d T^*} &=& \tau_{\mathrm{c,2}}^{-1} [ (G_2 - \alpha_2)
			  E_2 + K E_1'] + i \omega_2 E_2 +
			  \sqrt{\epsilon_2} \xi_2 (T^*),
		          \nonumber \\
  {d G_2\over d T^*} &=& \tau_{\mathrm{f,2}}^{-1} (p_2 - G_2 -
			 G_2 |E_2|^2).
\label{model}
\end{eqnarray}
Here $T^*$ is time, and
for $j=1,2,$ $E_j$ is the complex electric field, $G_j$ is
the gain, $\tau_{\mathrm{c},j}$ is
the cavity round-trip time, $\tau_{\mathrm{f},j}$ is the
fluorescence time of the lasing
ions, $p_j$ is the pump coefficient, $\alpha_j$ is
the cavity loss coefficient, $\omega_j$ is the
detuning of the laser from some common cavity mode,
and $K$ is a complex coupling coefficient with $K E_j'$
representing the degree of overlap of the
two lasers, with possible attenuation and dispersion taken into
account.  The noise term
$\sqrt{\epsilon_j} \xi_j (T^*)$ models
spontaneous emission, but for simplicity, we will consider only the
noiseless case $\epsilon_j = 0.$ Also, we will assume the media are
linear, nonabsorbing and nondispersive and the coupling is dissipative
so that $E_j' = E_j$ and $K$ is real.

The parameter $\mu$ in the first equation above is a symmetry-breaking
coefficient measuring the
extent of the feedback from the second laser to the first.
The case of symmetric coupling
($\mu=1, \tau_{\mathrm{c,1}}=\tau_{\mathrm{c,2}},
\tau_{\mathrm{f,1}}=\tau_{\mathrm{f,2}}$)
has been analyzed in \cite{TMRCLE97}.
In this paper, we focus instead on the case of
unidirectional coupling, i.e., $\mu = 0.$
When $\mu=0,$ the first laser is unaffected by the second,
and hence we may regard it as a driver.
Assuming that this driving laser is pumped above its lasing threshold
($p_1 > \alpha_1$), it is easy to show that its amplitude and phase
velocity settle down to constant values.  Specifically, the long-term 
state of the drive is given by
$E_1 = \sqrt{p_1 / \alpha_1 - 1} \exp \bbox{(} i (\Phi_{10} +
\omega_1 T^*) \bbox{)},$ where $\Phi_{10}$ is an arbitrary constant,
with constant gain $G_1 = \alpha_1$.

We assume these forms for $E_1$ and $G_1$ in the rest of the analysis.
Thus, although we have formulated our study in terms of one solid-state
laser driving another, the arguments and results we present hold for
more general situations, such as a solid-state laser subjected to
optical injection by other sources.

In a typical experimental setup using Nd:YAG lasers
\cite{TMRCLE97}, $\tau_{\mathrm{f},j}$ and
$\tau_{\mathrm{c},j}$ are both positive and are of the
orders $10^{-4}$ s and $10^{-10}$ s respectively. Thus we have
two vastly different time scales in the system. The detuning
$\omega_1 - \omega_2$ and the coupling $K$ are control parameters.
The detuning
has values typically of order $10^{5}$ Hz, while the coupling can be
varied over several orders of magnitude. In the symmetrically coupled
system \cite{TMRCLE97}, values of $K$ ranging from $O(10^{-8})$ to
$O(10^{-2})$ have
been used.  The pump $p_j$ and the loss $\alpha_j$ are both positive,
$O(10^{-2}),$
and their ratio $p_j / \alpha_j$ is typically an $O(1)$ quantity.


\section{Scaling and Symmetries}
\label{section_Nondim_Sym}

We scale the equations governing the response laser
by introducing the following dimensionless quantities:
\[
   \sigma = {\tau_{\mathrm{c,2}} \over \alpha_2 \tau_{\mathrm{f,2}}},
	    \
   B      = {p_2 \over \alpha_2}, \
   \tau   = {\alpha_2 T^* \over \tau_{\mathrm{c,2}}}, \
   F_2    = {G_2 \over \alpha_2}.
\]
Here $\sigma$ is a stiffness parameter, typically $O(10^{-4})$ in
experiments, characterizing the vast difference in the time scales of
the cavity round trip and fluorescence times in the response laser.
The smallness of $\sigma$ will be important in the subsequent
analysis.  The parameter $B$ is the dimensionless pump strength of the
response laser; it often plays the role of a control parameter in what
follows.  The variables $\tau$ and $F_2$ represent dimensionless time
and gain, respectively.

Next we change variables by going into a reference frame rotating
with the driver.   Let $X_2 \geq 0$ be the amplitude of the complex
field $E_2$, defined by
\[
E_2 = X_2 e^{i \Phi_2},
\]
and define $\Phi$ by
$\Phi = \Phi_2 - \Phi_1,$ where $\Phi_1$ is the phase of the driving
laser and $\Phi \sim
\Phi_2 - \Phi_{10} - ( \omega_1 \tau_{\mathrm{c,2}} / \alpha_2 ) \tau$
for sufficiently large $\tau.$
Assuming \cite{rectangular_cylindrical} $X_2 \neq 0,$ we obtain the
reduced system
\begin{mathletters}
\label{receiver}
  \begin{equation}
    {d X_2 \over d \tau} = (F_2 - 1) X_2 + k \cos \Phi
    \label{receiver:X2}
  \end{equation}
  \begin{equation}
    {d \Phi \over d \tau} = \Delta - {k \over X_2} \sin\Phi
    \label{receiver:Phi}
  \end{equation}
  \begin{equation}
    {d F_2 \over d \tau} = \sigma (B - F_2 - F_2 X_2^2),
    \label{receiver:F2}
  \end{equation}
\end{mathletters}
where
\begin{equation}
   \Delta = {(\omega_2 - \omega_1) \tau_{\mathrm{c,2}} \over \alpha_2},
            \;
   \ k = {K \sqrt{p_1 / \alpha_1 - 1} \over \alpha_2}.
\label{Delta_and_k}
\end{equation}
Here, $\Delta$ is a dimensionless measure of the frequency detuning of
the two lasers, and $k$ can be interpreted as either a dimensionless
coupling strength or injection amplitude.

By choosing the phase difference $\Phi$ as a variable, we have
eliminated the explicit time dependence in the original system by
rotating with the phase of the driver.  In this rotating reference
frame, a steady state now means a state in which the phase difference
between the two lasers, and not the phase $\Phi_2$ of the second laser
itself, is constant.  Such a state is said to be phase-locked.  The
particular case in which $\Phi=0$ (which is possible if and only if
$\Delta=0$) is called the coherent or in-phase state.

The analysis of the reduced system~(\ref{receiver})
will occupy most of this article.
As Eq.\ (\ref{receiver}) is invariant under
\begin{equation}
  \Delta \rightarrow - \Delta, \ \Phi \rightarrow - \Phi,
\label{symmetry:Delta}
\end{equation}
as well as under
\begin{equation}
  k \rightarrow -k, \ \Phi \rightarrow \pi + \Phi,
\label{symmetry:k}
\end{equation}
we will assume from now on that $k,\Delta \geq 0.$  Also,
$\sigma, B \geq 0$ by definition.

There is a slight catch that one should be aware of.  Although we
can assume $k \geq 0$ without any loss of mathematical generality,
there can still be physical consequences.  For example,
(\ref{symmetry:k}) allows us to change the sign of $k$, but at the
cost of transforming an in-phase solution to an antiphase one. In
fact, for certain systems of coupled lasers, $k$ can be
negative \cite{FCRL93}.


\section{Special Cases}
\label{section_Special_Cases}

The special case $k=0$ with general $\Delta$ is trivial: the driver
and the response laser are decoupled.  The reduced system
(\ref{receiver}) has a global attractor which is typically a
periodic orbit, corresponding to phase drift between the laser and
the drive.  In degenerate cases, the attractor can be a fixed point
for (\ref{receiver}), e.g., if there is also no detuning, or if the
pump for the response laser lies below the lasing threshold.

A more important special case is that of zero detuning:
$\Delta = 0$ with $k \neq 0.$
Then $\Phi=0$ (in-phase) and $\Phi=\pi$ (antiphase)
are invariant manifolds \cite{GH85}, the former attracting all initial
conditions
except those on the latter, which is repelling.  Hence, so
far as long-time behaviors are concerned, we can confine our
attention to the flow restricted to the manifold $\Phi=0.$

On $\Phi=0,$ Eq.\ (\ref{receiver})
simplifies to the planar vector field
\begin{eqnarray}
  {d X_2 \over d \tau} &=& (F_2 - 1) X_2 + k, \nonumber \\
  {d F_2 \over d \tau} &=& \sigma (B - F_2 - F_2 X_2^2).
\label{receiverPhi0}
\end{eqnarray}
The nullclines \cite{Strogatz94} are
\begin{mathletters}
\label{receiverPhi0nullcline}
  \begin{equation}
    F_2 = 1 - {k \over X_2}
    \label{receiverPhi0nullcline:a}
  \end{equation}
  \begin{equation}
    F_2 = {B \over 1 + X_2^2}.
    \label{receiverPhi0nullcline:b}
  \end{equation}
\end{mathletters}
Since Eq.\ (\ref{receiverPhi0nullcline:a}) represents an increasing
function of $X_2$ while Eq.\ (\ref{receiverPhi0nullcline:b})
represents a decreasing one, there can be at most one real
intersection. On the other hand, we may combine the two equations into
a cubic polynomial equation in $F_2$:
\[ F_2 [(F_2 - 1)^2 + k^2] - B (F_2 - 1)^2 = 0 \]
and see that there is at least one real solution in the interval
$(0,\min(1,B))$, by the intermediate value theorem. Hence there is
precisely one solution for $F_2$ in $(0,\min(1,B)),$ corresponding to
a unique fixed point.  An application of Dulac's criterion and the
Poincar\'{e}-Bendixson theorem \cite{Strogatz94}
to Eq.\ (\ref{receiverPhi0})
then reveals that this fixed point is actually globally stable
with respect to perturbations on the $X_2$$F_2$ plane.
Hence, in physical terms,
a globally stable in-phase state exists if and only if
there is no relative detuning between the two lasers.

Having exhausted the possibilities for these special cases,
we will assume $k,\Delta \neq 0$ from now on.


\section{Locked States}
\label{section_Locked_States}

\subsection{Existence and stability for general parameter values}

A locked state of the laser system corresponds to a fixed point of
Eq.\ (\ref{receiver}), which must satisfy
\begin{eqnarray}
  (F_2 - 1) X_2 + k \cos \Phi = 0, \nonumber \\
  \Delta - {k \over X_2} \sin\Phi = 0, \nonumber \\
  B - F_2 - F_2 X_2^2 = 0.
\label{fixedpoint}
\end{eqnarray}
In the limiting case $B=0,$ (i.e., the response laser is not pumped),
Eq.\ (\ref{fixedpoint}) can be readily solved to yield
\[
   F_2 = 0, \ \ \
   X_2 = {k \over \sqrt{1 + \Delta^2}}, \ \ \
   \Phi = \tan^{-1}\Delta.
\]
The eigenvalues of the Jacobian at this fixed point are
\[
   \lambda_1=-\sigma (1 + {k^2 \over {1 + \Delta^2}}), \ \ \
  \lambda_2, \lambda_3 = -1 \pm i \Delta,
\]
all having negative real parts.  This means, as $\Phi$ is the phase
difference between the two lasers, that the response laser is
passively driven into stable periodic motion
\[
   E_2 = {k \over \sqrt{1 + \Delta^2}} \exp \bbox{(} {i
	 (\tan^{-1}\Delta + \Phi_{10} + {\omega_1 \tau_{\mathrm{c,2}}
	 \over \alpha_2} \tau)} \bbox{)}
\]
with a constant phase-lead $\tan^{-1}\Delta$ ahead of the driver, even
though it is not pumped.

For $B>0$ so that $F_2 \neq 0$ at a fixed point, we may
eliminate $\Phi$ from Eq.\ (\ref{fixedpoint}) and get
\begin{mathletters}
\label{fixed_point_eqn}
  \begin{equation}
    X_2^2 = {B \over F_2} - 1,
    \label{fixed_point_eqn_a}
  \end{equation}
  \begin{equation}
    X_2^2 = {k^2 \over (F_2-1)^2 + \Delta^2}.
    \label{fixed_point_eqn_b}
  \end{equation}
\end{mathletters}
Consider the graphs of $X_2$ vs $F_2$ corresponding to
Eqs.~(\ref{fixed_point_eqn_a},\ref{fixed_point_eqn_b}).
The intermediate value theorem reveals that
\begin{itemize}
\begin{item}
  if $0< B \leq 1,$ there is only one fixed point, with
  $F_2 \in (0,B) \subseteq (0,1);$
\end{item}
\begin{item}
  if $1 < B < B_c,$ depending on parameter values, there
  may be one or three fixed points. There is always one in the interval
  $F_2 \in (0,1),$ and there may be a pair in $F_2 \in (1,B);$
\end{item}
\begin{item}
  if $B > B_c,$ again depending on parameter values,
  there may be one or three
  fixed points, all of which lie in $F_2 \in (1,B);$
\end{item}
\end{itemize}
where
\begin{equation}
  B_c \equiv 1 + {k^2 \over \Delta^2}
\label{B_c}
\end{equation}
represents a special value for the dimensionless pump strength.  In
Sec.\ \ref{subsection_Location_and_Stability}
we will show that $B_c,$ with a small correction,
is the critical value at which locked states lose stability.

To analyze the local stabilities of these fixed points, we apply
the Routh-Hurwitz criteria \cite{Uspensky48:p304,Hahn67:p19} to the
characteristic equation of Eq.\ (\ref{receiver}), and deduce
that the fixed point is locally stable if and only if
\begin{eqnarray}
  H_1 (F_2)    &:=& (B+2) F_2^2 - 2 F_2^3 - B (1 + \Delta^2) < 0,
	            \nonumber \\
  H_2 (F_2)    &:=& 2 \sigma (F_2^2 - B) - F_2 [(F_2 - 1)^2 + \Delta^2]
		    < 0, \nonumber \\
  H_{2'} (F_2) &:=& 2 F_2 (F_2 - 1) - \sigma B < 0,
	            \nonumber \\
  H_3 (F_2)    &:=& 2 (F_2 - 1)^3 + 2 \Delta^2 (F_2 - 1) \nonumber \\
	       &  & - {2 \sigma B \over F_2} (F_2 - 1) (F_2 - 2) -
	            2 \sigma F_2 (F_2 - 1) \nonumber \\
	       &  & + 2 \sigma^2 B - {2 \sigma^2 B^2 \over F_2^2} < 0,
\label{RouthHurwitz}
\end{eqnarray}
where $F_2$ is evaluated at the fixed point in question.  When
$H_1, H_3 < 0,$ the two inequalities $H_2 < 0$ and $H_{2'} < 0$
are equivalent and so we only need to check one of them.

For $B < B_c,$ we can verify by direct substitution
into (\ref{RouthHurwitz}) that the fixed point with
$F_2 < 1$ is stable.  As we increase $B$ through $B_c,$ this fixed
point crosses $F_2=1.$
The stabilities for the fixed points with $F_2 > 1$ are not so clear.
In fact, at $B=B_c$ we can solve the fixed-point equation
(\ref{fixed_point_eqn}) exactly and find that either $F_2 = 1$ or
\[
    F_2 =1 + {k^2 \over 2 \Delta^2} \pm \sqrt{ {k^4 \over 4 \Delta^4}
          - \Delta^2 - k^2},
\]
where the last pair exist if and only if $k^4 / 4 \Delta^4 \geq
\Delta^2 + k^2.$
Direct substitution into (\ref{RouthHurwitz}) reveals that
the fixed point with $F_2 = 1$ is stable (while the stabilities of the
other two fixed points remain unclear)
and hence there is no bifurcation at $B=B_c.$
Hence, the condition $B < B_c$ is sufficient but not necessary
for a stable locked state to exist.  In fact, for $\sigma = O(1),$
we have found numerical examples of stable locked states even when
$B > B_c$.  However, these examples are not of much physical interest,
given that $\sigma \ll 1$ for real lasers.  In the next subsection, we
will show that for $\sigma \ll 1$, the locking condition $B<B_c$
is indeed tight, with a correction term that approaches
zero as $\sigma$ does.

\subsection{Location and stability of locked states for $\sigma \ll 1$}
\label{subsection_Location_and_Stability}

From now on we will assume $\sigma \ll 1$.
For definiteness, consider the
scaling $k, \Delta = O(\sigma^a),$
and write
\[
k = \kappa \sigma^a, \; \;  \Delta = \delta \sigma^a.
\]
For $a > 0,$
the graph of Eq.\ (\ref{fixed_point_eqn_b}) has a narrow peak,
with $O(\sigma^a)$ width, at $F_2 = 1,$ and we can show that
there are three fixed points of Eq.\ (\ref{receiver}) if $B \leq B_c,$
while there is only one
fixed point if $B > B_c$ and $B - B_c \gg \sigma^{2a}.$
As we increase $B,$
somewhere in the region $B_c < B \leq B_c + O(\sigma^{2a}),$ two of the
fixed points collide.
To better understand what is going on, we will employ the Routh-Hurwitz
criteria (\ref{RouthHurwitz}).

For $B>1,$
we can find the location of the fixed points perturbatively if $a>0.$
There is always a fixed point with
\begin{equation}
  F_2 = F_{0}^* \equiv B - {\sigma^{2a} \kappa^2 B \over (B-1)^2} +
                        O(\sigma^{4a}).
\label{fixed_point_1}
\end{equation}
For $1 < B \leq B_c,$ there are a pair of fixed points
\cite{no_collision}
with
\begin{equation}
  F_2 = F_{\pm}^* \equiv 1 \pm \sigma^a \sqrt{ {\kappa^2 \over (B-1)}
                         - \delta^2 } + O(\sigma^{2a}).
\label{fixed_point_23}
\end{equation}
Direct substitution into (\ref{RouthHurwitz}) shows
that only the fixed point with
$F_2 = F_{-}^* $ is stable.
So we will concentrate on this fixed point and study the
mechanism through which it loses stability.
We expect that it will undergo a bifurcation in $B_c < B \leq B_c +
O(\sigma^{2a}).$

Bifurcation occurs if one of the functions in (\ref{RouthHurwitz})
vanishes \cite{Routh_Hurwitz_bifn}.  Specifically, the Routh-Hurwitz
condition $H_1 = 0$ corresponds to a zero eigenvalue, and is therefore
generically associated with a saddle-node bifurcation, whereas
$H_3 = 0$ indicates pure imaginary eigenvalues and a Hopf bifurcation.
(These results follow immediately from the observation that the
characteristic polynomial is cubic with real coefficients.)

We solve (\ref{RouthHurwitz}) in
conjunction with the fixed-point equation (\ref{fixed_point_eqn}) by
the method of dominant balance \cite{Zwillinger92}.  
Let $B = B_c + \varepsilon.$  Then $H_1 (F_2) = 0$ if
\begin{eqnarray}
  \varepsilon &=& \varepsilon_1
	      \equiv {\sigma^{2a} \delta^2 B_c^2 \over 4 (B_c - 1)} +
	             O(\sigma^{4a}), \nonumber \\
  F_2 &=& 1 + {\sigma^{2a} \delta^2 B_c \over 2 (B_c - 1)} +
	  O(\sigma^{4a}),
\label{RH1}
\end{eqnarray}
and $H_3 (F_2) = 0$ if
\begin{eqnarray}
  \varepsilon &=& \varepsilon_3
	      \equiv \sigma^{2-2a} B_c^2 (B_c - 1) /
		     \delta^2 + O(\sigma^{3-4a}),
		     \nonumber \\
  F_2 &=& 1 + \sigma^{2-2a} B_c (B_c - 1) / \delta^2 +
          O(\sigma^{3-4a}) \nonumber \\
      & & {\mathrm \ \ for \ \ } a < 1/2
\label{RH3}
\end{eqnarray}
There is no self-consistent solution, with $0 < \varepsilon \ll 1,$ for
$H_3 (F_2) = 0$ if $a > 1/2.$
If $a < 1/2,$ then $\varepsilon_3 \ll \varepsilon_1$ and so $H_3$
becomes zero before $H_1$ does, indicating a Hopf bifurcation.
(Whether it is sub- or supercritical is undetermined at this stage in
the calculation.  But
see Sec.\ \ref{subsection_PeriodicMotion_Hopf_Discussion}.)
If $a > 1/2,$ then $H_1$ becomes
zero at $\varepsilon = \varepsilon_1,$
indicating a zero-eigenvalue bifurcation, which, given the absence of
any appropriate symmetry,
is expected to be a saddle-node bifurcation \cite{GH85}.
We have confirmed these bifurcation scenarios numerically, using the
bifurcation package \textsc{auto} \cite{DWF96}.

The argument above leaves a gap at $a=1/2,$ in which case $H_1$ and
$H_3$ both vanish when $\varepsilon = O(\sigma),$ corresponding to a
bifurcation of higher codimension, and potentially leading to a 
complicated outcome.  We will look into this case separately in
Sec.\ \ref{section_Beyond_Periodicity}.
But as far as the fixed point is concerned, it loses stability at
$B = B_c + O(\sigma).$  Hence, no matter whether the scaling exponent
$a$ is greater than, less than, or equal to 1/2, we can conclude that
$B_c$ is a tight estimate of the unlocking threshold, with an error
at most of $O(\sigma).$

To sum up, for $\sigma \ll 1$ we have shown that:
\begin{itemize}
\begin{item}
  if $B < B_c + \varepsilon,$ there is one locally stable fixed
  point.
  This corresponds to a stable injection-locked state, with a nonzero
  phase difference unless there is no detuning.
\end{item}
\begin{item}
  if $B > B_c + \varepsilon,$ then there may be one or three fixed
  points, all in the region $F_2 \in (1,B),$ and they are all saddles.
 Hence there is no stable locked state in this case.
\end{item}
\end{itemize}
Here, $0 < \varepsilon = \min (\varepsilon_1, \varepsilon_3) \leq
O(\sigma),$ with  $\varepsilon_1, \varepsilon_3$ given by
Eqs.\ (\ref{RH1}) and (\ref{RH3}), and $B = B_c + \varepsilon$ is the
parameter value at which the unlocking transition occurs.

Taking advantage of the symmetries (\ref{symmetry:Delta}) and
(\ref{symmetry:k}), we may rewrite the locking condition $B < B_c +
\varepsilon$
in terms of the physical parameters defined in Eq.\ (\ref{model}):
\begin{equation}
  |K| \sqrt{ {p_1 \over \alpha_1} - 1} >
      | \omega_1 - \omega_2 | \tau_{\mathrm{c,2}}
      \sqrt{ {p_2 \over \alpha_2} - 1 - \varepsilon},
\label{lock_criterion}
\end{equation}
with $0 < \varepsilon \leq O(\tau_{\mathrm{c,2}} / \alpha_2
\tau_{\mathrm{f,2}}).$  In physical terms, stable locking occurs if the
coupling $K$ and the injection intensity (controlled by the pump
strength $p_1$) are sufficiently strong to overcome the relative
detuning between the drive and the response laser.

It is also noteworthy that the right-hand side of the locking condition
depends on the pump strength $p_2$ of the response laser.  In
applications involving injection locking, a laser with a low power
output but very accurate frequency is often used to control a strong
but sloppy one.  The criterion
(\ref{lock_criterion}) indicates that locking is more difficult if the
response laser has a stronger pump $p_2.$

\subsection{Discussion}

The condition for locking is commonly derived in the limit of small
injection intensity (which equals $p_1 / \alpha_1 - 1$ in our
notation).  This derivation relies on the slowly varying envelope
approximation, along with
the further assumptions that the intensity of the response laser
is at the free-running output level and that
the gain has saturated.  Then the system of governing equations is
reduced to a phase model equivalent to Adler's equation
\cite{Siegman86} \cite{Barillet96}.
The traditional result for the locking condition is, in terms of our
symbols, $ \Delta < {k / \sqrt{I_2}} $ where $I_2$ is the free-running
intensity of the driven laser.  Since $I_2 = B-1,$ this condition
reduces to our condition $B < B_c.$

We have re-derived this well known result in a more general context.
In particular, we have shown that a sufficient condition for locking
is $B < B_c$, even when $k$ is {\it not} small and the phase model no
longer holds.   [Recall that $k$ is the dimensionless product of the
coupling strength and the injection amplitude; see
Eq.~(\ref{Delta_and_k}).] Moreover, our approach allows us to
calculate the correction terms
in the formulas for the locking threshold, namely Eqs.\ (\ref{RH1}) and
(\ref{RH3}),  in the limit of small $\sigma =\tau_{\mathrm{c,2}} /
(\alpha_2 \tau_{\mathrm{f,2}}).$

To see why a simple phase model can give the correct locking condition,
consider the least-stable eigendirection of the locked state.  Suppose
the coupling is weak, in the sense that
$k = O(\Delta) \ll \sigma^{1/2}$.  At the bifurcation point, the
soft mode (the eigenvector associated with the
zero eigenvalue) is given by
\[
   (X_2, \Phi, F_2) = (-\sigma^a, {2 (B_c - 1) \over \kappa B_c},
		      {2 \sigma^a \kappa \over \delta B_c}) +
		      O(\sigma^{2a}).
\]
For small $\sigma$, the first and third components of this eigenvector
are small, indicating that the soft mode lies nearly along the phase
direction.  Hence when the coupling is weak, the phase direction alone
determines the locking condition, to a good approximation.

More generally, one can ask ``When can the full model
be replaced by a phase model?''.  We will show in the following section
that this reduction is valid when $k$ and  $\Delta$ are small.


\section{Weak Coupling}
\label{section_Weak_Coupling}

If $B > B_c + \varepsilon,$ there is no stable fixed point of Eq.\
(\ref{receiver}), so the response laser cannot lock to the drive.  The
desynchronized dynamics then strongly depend on the relative sizes of
the parameters $k, \Delta$ and $\sigma.$  For definiteness we will
continue to assume that $\sigma \ll 1$ and $k, \Delta = O(\sigma^a).$

In this section, we focus on the weak coupling case $a > 1/2$, and show
that unlocking occurs via a saddle-node infinite-period
bifurcation.  All solutions are then attracted to a stable periodic
solution for the reduced system; this periodic solution corresponds
physically to a phase-drifting state.

Before we go on, a warning is in order: we will consider only lowest
order effects in this section for simplicity.  In particular, we
will not pick up the $\varepsilon$ corrections [Eqs.\ (\ref{RH1}) and
(\ref{RH3})] for the locking condition.

Suppose that $k = O(\Delta) \ll \sigma^{1/2},$ or equivalently,
\[ |K| \sqrt{{p_1 \over \alpha_1} - 1} \sim
   | \omega_1 - \omega_2 | \tau_{\mathrm{c,2}} \ll
   \sqrt{{\alpha_2 \tau_{\mathrm{c,2}} \over \tau_{\mathrm{f,2}}}}
\]
in terms of the physical parameters in Eq.\ (\ref{model}).
In this limit, we can reduce the full model Eq.\ (\ref{receiver})
to a phase model by the following argument.  Let $u = F_2 - 1$ and
introduce the self-consistent assumption $X_2 = O(1)$ (which we
find to be justified by numerics).  Then we observe that if
$u \gg k,$ the term $k \cos\Phi$ in Eq.\ (\ref{receiver:X2})
is negligible and so the $X_2,u$ dynamics decouples from that of
$\Phi$ and constitutes a two-dimensional system.  Next, by a phase
plane analysis and diagonalizing the Jacobian we can show that
$X_2 \rightarrow \sqrt{B-1}$ and $u$ decreases toward 0,
with an $O(\sigma^{1/2})$ spiralling frequency and an
$O(\sigma)$ relaxation rate.
So after an $O(\sigma^{-1})$ transient time, $u$ becomes
comparable to $k$ and we can no longer neglect the term $k \cos\Phi$
in Eq.\ (\ref{receiver:X2}).

As $X_2 = O(1),$ Eq.\ (\ref{receiver:Phi}) indicates that $\Phi$
evolves with a characteristic rate of size $O(\Delta,k)$ .  If this
characteristic rate is much smaller than that for the spiralling
$X_2,u$ dynamics, which is true if $\Delta,k \ll \sigma^{1/2},$ we can
take $k \cos\Phi$ to be a constant with an $O(k^2)$ error, and deduce
that $X_2,u$ relax to their equilibrium values with an
$O(\sigma^{1/2})$ characteristic rate.

Assuming this relaxation has happened,
Eq.\ (\ref{receiver:Phi}) becomes
\begin{equation}
   {d \Phi \over d \tau} = \Delta - {k \over \sqrt{B-1}} \sin\Phi +
			   O(k^2),
\label{Adler}
\end{equation}
which is recognized as Adler's equation [Eq.(\ref{PLL})].
The important feature is that $\Phi$ either
approaches a constant or is strictly increasing, with
$\Phi {\mathrm \ mod \ } 2 \pi$ being periodic with period
$2 \pi / \sqrt{\Delta^2 - k^2 / (B-1)},$  depending on whether $B$
is smaller than or larger than $1+k^2/\Delta^2,$ with an $O(k^2)$
error.

So, for large $\tau,$ the unlocked dynamics are given by
\begin{eqnarray}
  X_2 &\sim& \sqrt{B-1} + {k B \cos\Phi(\tau) \over 2 (B-1)} + O(k^2),
                    \nonumber \\
  F_2 &\sim& 1 - {k \cos\Phi (\tau) \over \sqrt{B-1}} + O(k^2),
\label{phase_model_soln}
\end{eqnarray}
where $\Phi (\tau)$ is governed by Eq.\ (\ref{Adler}).
Hence $X_2 (\tau)$ and $F_2 (\tau)$ oscillate periodically, with $O(k)$
amplitudes and the same Fourier spectra as $\cos \Phi.$, up to overall
multiplicative constants.   This last feature
supplements the observation, noted in \cite[p.~1153]{Siegman86}, that
the laser intensity ``acquires distortion at higher harmonics'' as
the locking threshold is approached.  In fact, by insisting on a phase
model,
Siegman \cite[p.~1148]{Siegman86} analyzed the dynamics of a laser with
external injection outside the locking regime.  Now we have justified
this insistence by showing that in an appropriate regime in the
parameter space, namely $k = O(\Delta) \ll \sigma^{1/2}$, the laser
system indeed approaches a state with $X_2 \simeq \sqrt{B-1}, F_2
\simeq 1,$ starting from arbitrary initial conditions.

The saddle-node bifurcation found in
Sec.\ \ref{subsection_Location_and_Stability} for $a > 1/2$ can now be
identified more specifically as a saddle-node infinite-period
bifurcation \cite{Strogatz94}.  For $B > B_c + \varepsilon,$
the phase increases strictly with a nonuniform speed, with a
bottleneck near $\Phi = \sin^{-1} (\Delta \sqrt{B-1} / k).$
These behaviors are illustrated in
Fig.\ \ref{fig_saddle_node}.

The argument above clearly does not hold if $k, \Delta =
O(\sigma^{1/2}),$ as then the phase dynamics has a characteristic rate
comparable to that of the spiralling motion on the $u X_2$
cross-sections.  We will address this case in
Sec.\ \ref{section_Beyond_Periodicity}.


\section{Strong Coupling}
\label{section_Strong_Coupling}

We turn now to the case of strong coupling: $k = O(\Delta)  \gg
\sigma^{1/2}.$   The main result is that unlocking occurs via a
supercritical Hopf bifurcation, leading to a globally attracting limit
cycle for Eq.\ (\ref{receiver}) when $B > B_c + \varepsilon.$  We will
investigate in Sec.\ \ref{subsection_Two_Timing} the case
$k = O(\Delta) = O(1).$  Then in Sec.\ \ref{subsection_Rescaling} we
use a rescaling argument to show that this scenario also subsumes the
ostensibly more general case $k = O(\Delta) \gg \sigma^{1/2}$.

\subsection{Two-timing calculations}
\label{subsection_Two_Timing}

Assume that $\sigma \ll 1$, and  $k,\Delta=O(1)$.  By defining
\[ E = X_2 e^{i \Phi}, \]
which represents the complex electric field of the response laser as
observed in a frame rotating with the driver, we can combine
Eqs.\ (\ref{receiver:X2}) and (\ref{receiver:Phi}) into
\begin{equation}
  { d E \over d \tau} =
    [(F_2 - 1) + i \Delta] E + k.
\label{complex_eqn}
\end{equation}
For $\sigma \ll 1,$ $F_2$ is slowly varying and so may be taken to be a
constant except over long time scales of order $O(1/\sigma)$.
If $F_2$ were really a constant, the exact solution to
Eq.\ (\ref{complex_eqn})
could be found as
\begin{equation}
E = E_c \exp \bbox{(} [(F_2-1)+ i \Delta] \tau \bbox{)} -
    {k \over (F_2-1) + i \Delta},
\label{ifexact}
\end{equation}
where $E_c$ is a complex constant. Hence
$X_2 = |E|$ would decay exponentially to
${k / \sqrt{(F_2-1)^2 + \Delta^2}}$ if $F_2<1$
and grow exponentially to infinity if $F_2>1.$
In the first scenario where $F_2 < 1,$ we have
\[
         {d F_2 \over d \tau}
  \simeq \sigma [B - F_2 (1 + {k^2 \over (F_2-1)^2 + \Delta^2})] > 0
\]
if $B > B_c$ and  $F_2 \simeq 1.$
So $F_2$ would grow for $F_2 < 1.$ In the second case where $F_2 > 1,$
${d F_2 / d \tau} \simeq - \sigma F_2 X_2^2 < 0$ and so $F_2$ would
decay.  Hence we expect that, after initial transient instabilities
for $\tau \ll O({1 / \sigma}),$ the system will settle down to its
eventual fate, with $F_2 \simeq 1$ and
$E \simeq E_c \exp(i \Delta \tau) + i k / \Delta$ for all sufficiently
large $\tau,$ and this is indeed what we find in our numerics
(not shown).

Now we investigate the structure of the attractor of
Eq.\ (\ref{receiver}) by two-timing \cite{Zwillinger92}.
We first define a slow time $T=\sigma \tau.$ Abusing notation, we have
\[
  {d \over d\tau} f (\tau,T)
     = {\partial f \over \partial \tau} +
	 \sigma {\partial f \over \partial T},
\]
where $f$ denotes either $F_2$ or  $E.$
Assuming transients have already decayed, we let
$F_2 = 1 + \sigma F_2^{(1)} + O(\sigma^2)$ and
$E = E^{(0)} + \sigma E^{(1)} + O(\sigma^2).$
Then we find
\begin{eqnarray*}
  E^{(0)}   &=& A^{(0)} (T) e^{i [ \theta^{(0)} (T) +
	        \Delta \tau ]} + {i k \over \Delta}, \\
  F_2^{(1)} &=& f(T) + \{ B - 1 - [A^{(0)} (T)]^2 -
		{k^2 \over \Delta^2} \} \tau \\
            & &	+ {2k \over \Delta^2} A^{(0)} (T)
		\cos \bbox{(} \theta^{(0)} (T)+
	        \Delta \tau \bbox{)}.
\end{eqnarray*}

To remove the secularity so that $F_2^{(1)}$ remains bounded
(which is necessary for $F_2$ to stay close to $1$), we need
\begin{equation}
  A^{(0)} = \sqrt{B - (1+{k^2 \over \Delta^2})} = \sqrt{B-B_c}.
\label{Hopf_amplitude}
\end{equation}
This is well defined if and only if $B \geq B_c,$ which
is, to lowest order,
the condition for the fixed point of (\ref{receiver})
to be unstable (Sec.\ \ref{section_Locked_States}).
Substituting the results for $E^{(0)}$ and $F_2^{(1)}$ into the
$E^{(1)}$ equation and suppressing secularity so that it is possible
to have $E \simeq E_c \exp(i \Delta \tau) + i k / \Delta,$ we get
\[
   f = 0, \ \ \
   \theta^{(0)} = {k^2 \over \Delta^3} T + \psi^{(0)}
	        = \sigma {k^2 \over \Delta^3} \tau + \psi^{(0)}
\]
for some constant $\psi^{(0)}.$ Hence, for time $\tau = O(1/\sigma),$
and assuming the system is \emph{already on the attractor,} we find
\begin{eqnarray}
  E &=& A^{(0)} \exp \bbox{(} i [ (\Delta +
	\sigma {k^2 \over \Delta^3} ) \tau +
	\psi^{(0)}] \bbox{)} + {i k \over \Delta} + O(\sigma)
        \nonumber \\
  F_2 &=&
     1 + 2 \sigma {k \over \Delta^2} A^{(0)}
     \cos \bbox{(} (\Delta+\sigma {k^2 \over \Delta^3})\tau +
     \psi^{(0)} \bbox{)} + O(\sigma^2), \nonumber \\
  \ \
\label{twotimingresults}
\end{eqnarray}
where $A^{(0)} = \sqrt{B - B_c}$ and $\psi^{(0)}$
is an arbitrary phase constant.  These results
have been compared with numerics and show good agreement.  Note that
the result is singular if $\Delta \rightarrow 0.$  A similar singular
limit has been observed in a model of a $\mathrm{CO_2}$ laser
\cite{SO94}.

\subsection{Discussion}
\label{subsection_PeriodicMotion_Hopf_Discussion}

\subsubsection{Physical interpretation}

An inspection of $E$ in Eq.\ (\ref{twotimingresults}) reveals
that
\begin{itemize}
\begin{item}
  if $A^{(0)} < {k / \Delta},$ the phase difference $\Phi$
  oscillates between
  ${\pi / 2} \pm \tan^{-1} ({\Delta A^{(0)} / k}).$  The response laser
  is said to be {\it phase-trapped} to the drive \cite{KCPMW82}:
  both lasers have \emph{the same average frequencies} but their
  relative phase varies periodically.  In other words, they are
  frequency-locked but not phase-locked.
\end{item}
\begin{item}
  if $A^{(0)} > {k / \Delta},$ then $\Phi$ increases monotonically at
  an almost uniform rate, corresponding to a state of phase drift.
\end{item}
\end{itemize}
These different behaviors, together with the regimes in parameter
space in which they occur, are depicted in
Fig.\ \ref{fig_hopf}.  This figure should be
compared with Fig.\ \ref{fig_saddle_node}, which shows that
for $k = O(\Delta) \ll \sigma^{1/2},$ $\Phi$ changes directly from
locking to
drifting at the critical pump value $B = B_c + (\sigma^{2a}),$
with no intervening possibility of phase trapping.

Figure~\ref{fig_order1} illustrates the accuracy of our analytical
approximations.  For a given value of $k$, we compute the time
series of the intensity $I = X_2^2$.  The local minima and maxima
of the intensity are plotted in Fig. \ref{fig_order1} as a function
of $k$.  For $k>2$, the intensity is constant because the system has
a stable fixed point corresponding to a locked state; hence only a
single-valued branch of data is seen.  As $k$ decreases through the
unlocking threshold, a limit cycle is born, causing two branches to
bifurcate continuously from of the locked state, as shown in
Fig. \ref{fig_order1}.  This splitting is what one would expect for
a supercritical Hopf bifurcation -- the intensity now oscillates
sinusoidally, so the lower branch corresponds to local minima of
the intensity time series, and the upper branch corresponds to maxima.
The curves passing through the data are analytical predictions given
by Eq.~(\ref{twotimingresults}).

While the solution (\ref{twotimingresults}) is periodic in the
reference frame rotating with the driver, back in the laboratory
frame the solution involves two frequencies.  The phase of the
receiver laser is then
\begin{eqnarray*}
   \Phi_2    &=& \Phi_1 + \Phi \\
	  &\sim& \Phi_{10} + {\omega_1 \tau_{\mathrm{c,2}} \over
		 \alpha_2 } \tau + \arg E
\end{eqnarray*}
and is in general quasiperiodic.

\subsubsection{Hopf bifurcation at $B = B_c + O(\sigma^2)$}

The calculations in Sec.\ \ref{subsection_Two_Timing}, and in
particular Eq.\ (\ref{Hopf_amplitude}), are correct only to the lowest
order in $\sigma.$ If we proceed to higher orders, we find
\[
   A = \sqrt{B - \left( B_c + {\sigma^2 B_c^2 (B_c - 1) \over \Delta^2}
       \right) } + O(\sigma^3).
\]
From this and Eq.\ (\ref{twotimingresults}) we see that the radius
of the stable limit cycle scales as $\sqrt{B - B_{\mathrm{bifn}}},$
where $B_{\mathrm{bifn}} = B_c + O(\sigma^2)$ is the bifurcation value 
at which the limit cycle is born [compare with Eq.\ (\ref{RH3})], and 
the frequency $\Delta + \sigma {k^2 / \Delta^3}$ is an $O(1)$ quantity.
These results strongly hint that this is a supercritical Hopf
bifurcation
\cite{Strogatz94}. Indeed, we can prove that this is the case
for $\sigma \ll 1$ and $k, \Delta = O(1)$, as follows.

For $B$ close to $B_{\mathrm{bifn}},$ we may take
$\varepsilon = B - B_{\mathrm{bifn}}$ as a small
parameter.  By perturbing first with respect to $\sigma$ and then with
respect to $\varepsilon,$ we find that the three eigenvalues at the
relevant fixed point are
\begin{eqnarray*}
  \lambda_1 &=& i (\Delta + {\sigma k^2 \over \Delta^3}) +
		\varepsilon [{1 \over B_c} -
		{\sigma k^2 \over \Delta^4 B_c}
		(1 - i \Delta)] + O(\sigma^2,\varepsilon^2), \\
  \lambda_2 &=& \overline{\lambda_1},  \\
  \lambda_3 &=& - \sigma B_c +
		  {2 \varepsilon \sigma k^2 \over \Delta^4 B_c} +
		  O(\sigma^2,\varepsilon^2),
\end{eqnarray*}
where an overbar denotes a complex conjugate.
As $\varepsilon$ goes through zero, $\lambda_1$ and $\lambda_2$
cross the imaginary axis with nonzero speed, while $\lambda_3$ remains
real and negative.  Thus, by Hopf's theorem \cite{HK76}, a limit cycle
is present on one side of the bifurcation point.  Moreover, this limit
cycle is centered at the fixed point, and has angular frequency given
by the imaginary part of the conjugate pair of the eigenvalues. All
these results are in agreement with Eq.\ (\ref{twotimingresults}),
and we have also confirmed them numerically.

To decide whether the bifurcation is sub- or supercritical, we
use the Poincar\'{e}-Lindstedt method \cite{Strogatz94}
to seek a periodic solution.
We find that such a solution, be it stable or not, can exist only
if $B > B_{\mathrm{bifn}},$  i.e., when the fixed point is unstable,
indicating that the bifurcation is supercritical.
The periodic solution thus obtained is, as expected, the same as
that given by Eq.\ (\ref{twotimingresults}).
However, while the Poincar\'{e}-Lindstedt
method tells us nothing about the stability of the solution, it has the
merit that it is uniformly valid for all time, thus confirming the
speculation, based on numerics and the two-timing analysis in
Sec.\ \ref{subsection_Two_Timing},
that there is indeed a periodic solution.

\subsection{Rescaling for
               $\sigma \ll 1$ with
		$\lowercase{k} = O(\Delta) \gg \sigma^{1/2}$}
\label{subsection_Rescaling}

Having considered the case $k, \Delta = O(1),$
we will now demonstrate that as long as
$k = O(\Delta) \gg O(\sigma^{1/2}),$
the lowest-order dynamics remain the same as before.  Thus, any
coupling and detuning that satisfy these milder bounds are still
sufficiently strong to preserve the qualitative features seen earlier.
In this sense, the condition $k = O(\Delta) \gg O(\sigma^{1/2})$
establishes the demarcation line for the regime of ``strong'' coupling
and detuning.

By defining $u$ via $F_2 = 1 + \sigma u,$ we may rewrite Eq.\
(\ref{receiver}) as
\begin{mathletters}
\label{receiver_scaled}
  \begin{equation}
    {d X_2 \over d\tau} = \sigma u X_2 + k \cos\Phi,
    \label{receiver_scaled:X2}
  \end{equation}
  \begin{equation}
    {d \Phi \over d\tau} = \Delta - {k \over X_2} \sin\Phi,
    \label{receiver_scaled:Phi}
  \end{equation}
  \begin{equation}
    {d u \over d\tau} = B - (1 + \sigma u) (1 + X_2^2).
    \label{receiver_scaled:F2}
  \end{equation}
\end{mathletters}
For $\Delta,k=O(1),$ we observe in numerics that
$F_2 = 1 + O(\sigma)$ so that $u=O(1).$
More generally, if $k, \Delta = O(\sigma^a),$ we may write
$\Delta = \delta \sigma^a, k = \kappa \sigma^a,$ with $\delta, k =
O(1),$ and rescale time by setting
$t = \sigma^a \tau.$ Then
Eq.\ (\ref{receiver_scaled}) becomes
\begin{mathletters}
\label{receiver_rescaled}
  \begin{equation}
    {d X_2 \over d t} = \sigma^{1-2a} u X_2 + \kappa \cos\Phi,
    \label{receiver_rescaled:X2}
  \end{equation}
  \begin{equation}
    {d \Phi \over d t} = \delta - {\kappa \over X_2} \sin\Phi,
    \label{receiver_rescaled:Phi}
  \end{equation}
  \begin{equation}
    {d u \over d t} = B - (1 + \sigma^{1-a} u) (1 + X_2^2).
    \label{receiver_rescaled:F2}
  \end{equation}
\end{mathletters}

For $a < 1/2,$ our numerical simulations indicate that
$F_2 = 1 + O(\sigma^{1-a})$ so that $u=O(1).$ Hence we can invoke a
self-consistent argument to identify, to lowest orders,
Eq.\ (\ref{receiver_rescaled}) with Eq.\ (\ref{receiver_scaled}),
with $\sigma^{1-2a}, t, \delta, \kappa$ replacing
$\sigma, \tau, \Delta, k$ respectively, since the terms
$\sigma u$ in Eq.\ (\ref{receiver_scaled:F2}) and $\sigma^{1-a} u$ in
Eq.\ (\ref{receiver_rescaled:F2}) do not affect the lowest order
phenomena and are therefore negligible.
As $\delta, \kappa = O(1),$ we can apply the
same two-timing analysis as in Sec.\ \ref{subsection_Two_Timing},
but with the time scales $t$ and $\sigma^{1-2a} t,$
and get results similar to those found earlier.
In terms of the variables used before rescaling, the lowest order
terms are found to be
\begin{eqnarray}
  X_2 e^{i \Phi} =
    E &=& A^{(0)} \exp \bbox{(} i [(\Delta+\sigma {k^2 \over \Delta^3})
          \tau + \psi^{(0)}] \bbox{)} \nonumber \\
      & & + {i k \over \Delta} + O(\sigma^{1-2a}), \nonumber \\
  F_2 &=& 1 + 2 \sigma {k \over \Delta^2} A^{(0)}
          \cos \bbox{(} (\Delta+\sigma {k^2 \over \Delta^3})\tau +
	  \psi^{(0)} \bbox{)} \nonumber \\
      & & + O(\sigma^{2-2a,3-5a}),
\label{hopf_rescaled_result}
\end{eqnarray}
where, as before,
$A^{(0)} = \sqrt{B - B_c} = \sqrt{B - 1 - k^2/\Delta^2}$ and
$\psi^{(0)}$ is an arbitrary phase constant.  This result is
the same as Eq.\ (\ref{twotimingresults}) to lowest order, exhibiting
periodic motion born out of a
supercritical Hopf bifurcation as $B$ increases through $B_c$ from
below.  One slight difference is that the next higher-order terms will
be $O(\sigma^{1-2a})$ instead of $O(\sigma).$ Also, corresponding
behaviors in this new system will occur on a slower
time scale, as
$t = \sigma^a \tau \ll \tau,$ if $a > 0.$

Clearly the argument above breaks down if $a=1/2,$ i.e., if
$k = O(\Delta) = O(\sigma^{1/2})$, as then $\sigma^{1-2a}$ in
Eq.\ (\ref{receiver_rescaled:X2}) is no
longer small as assumed by the
perturbative treatment. Another way of putting this is, the two time
scales, which are $O(1)$ and $O(\sigma^{-1+2a})$ in $t,$ collapse
into one as $a \rightarrow 1/2.$
In the next section, we turn our attention to this distinguished regime
$k = O(\Delta) = O(\sigma^{1/2})$. For symmetrically coupled lasers,
this is the  regime where subharmonic resonance, amplitude
instabilities, and chaos were discovered \cite{TMRCLE97,EBM87}.


\section{Intermediate Coupling}
\label{section_Beyond_Periodicity}

\subsection{Overview}
\label{subsection_Overview_for_sigma12}

The cases studied earlier allow us to anticipate some aspects of the
possible unlocking behavior for intermediate coupling.
From Secs.\ \ref{section_Locked_States}, \ref{section_Weak_Coupling},
and \ref{section_Strong_Coupling},
we know that for weak coupling $(a > 1/2)$, the fixed point
corresponding to the stable
locked state disappears in a saddle-node infinite-period bifurcation
when $B
- B_c = O(\sigma^{2a}),$ while for strong coupling $(a < 1/2)$, it
loses stability in a supercritical Hopf bifurcation when $B - B_c =
O(\sigma^{2-2a}).$  Hence, by sandwiching, we
expect that for intermediate coupling $(a=1/2)$, the stable fixed point
will lose
stability when $B - B_c = O(\sigma),$ probably in a codimension-two
bifurcation
combining the features of a saddle-node infinite-period bifurcation
and a supercritical Hopf bifurcation.

For $B$ slightly above
the bifurcation value, we expect the dynamics to be a combination of
the two cases studied previously.  The long-time
behavior should be a spiralling motion that slowly passes through the
bottleneck near
\[ X_2 = \sqrt{B-1}, \;  \;
   \Phi = \sin^{-1} ({\Delta \sqrt{B-1} \over k}), \;  \;
   F_2 = 1,
\]
together with a fast re-injection roughly along the circle
\[ X_2 = \sqrt{B-1},  \;  \;  F_2 = 1. \]
Meanwhile, there is a global relaxation towards this attractor.  The
interplay among these three mechanisms can lead to complicated
behavior, as we will see below.

In contrast, in the case of symmetrical coupling ($\mu=1$ in
Eq.\ (\ref{model})), the phase difference decouples from the intensity
and the gain,  if one also assumes that both lasers have equal gains
and intensities at all times \cite{TMRCLE97}.  This decoupling of the
phase dynamics leaves only two mechanisms: a global
relaxation toward the attractor, and a phase advancement
(re-injection) along the attractor.  In this sense, the
unidirectionally coupled
case might be prone to greater dynamical complexity.

\subsection{Numerics}

In the intermediate coupling regime $k = O(\Delta) = O(\sigma^{1/2}),$
Eq.\ (\ref{receiver_rescaled}) becomes
\begin{eqnarray}
   {d X_2 \over d t} &=& u X_2 + \kappa \cos \Phi,  \nonumber \\
   {d \Phi \over d t} &=& \delta - {\kappa \over X_2} \sin \Phi,
       \nonumber \\
   {d u \over d t} &=& B - (1 + \sigma^{1/2} u) (1 + X_2^2),
\label{receiver_sigma12}
\end{eqnarray}
 with $\kappa \sim \delta = O(1) \gg \sigma^{1/2}$.  In rectangular
coordinates, with
$x=X_2 \cos \Phi, y=X_2 \sin \Phi,$ this system can be rewritten as
\begin{eqnarray}
   {d x \over d t} &=& ux - \delta y + \kappa,  \nonumber \\
   {d y \over d t} &=& uy + \delta x, \nonumber \\
   {d u \over d t} &=& B - (1 + \sigma^{1/2} u) (1 + x^2 + y^2).
\label{blown_up_rectangular}
\end{eqnarray}

To probe the dynamics, we will vary $\kappa$ and hold the other
parameters fixed.  This may seem unnatural, given that we have been
using $B$ as a control parameter so far.  However, it turns out that
if we vary $\kappa$ instead of $B$, the features we wish to emphasize
will stand out more clearly.  In experiments, both parameters are
easily tunable.

Figure~\ref{fig_sigma05} illustrates the complicated dynamics
that occur in this system.  The plot is an orbit diagram, shown in
the same format as  Fig.~\ref{fig_order1}.  The local minima and
maxima of the intensity's time series are shown as the coupling
$\kappa$ increases toward the locking threshold.  For the particular
parameters chosen in the simulation, the locking threshold is at
$\kappa_c = 0.9986 = \delta \sqrt{B-1} + O(\sigma)$.  Of course,
for $\kappa > \kappa_c$, the results are simple:
there is a single branch of constant intensity, corresponding to the
locked state.  But for $\kappa < \kappa_c$ we see
an intricate pattern of local maxima and minima.

The most striking feature occurs for $0.83 \leq \kappa \leq \kappa_c,$
where the diagram consists of a sequence of similar patterns that are
scaled down as $\kappa$ increases.  Figure~\ref{fig_sigma05}(b)
shows the region close to $\kappa = \kappa_c$ in greater detail.  The
self-similar structure appears to persist all the way to the unlocking
threshold at $\kappa = \kappa_c$.  This scenario is not the standard
period-doubling route to chaos, nor any of the other familiar
bifurcation cascades.  We do not fully understand what is
happening here.  Some first steps toward understanding this remarkable
structure will be presented in
Sec.\ \ref{section_A_Singular_Limit}.

We are mainly interested in the self-similar structure near the
unlocking threshold, but some other features of
Fig.\ \ref{fig_sigma05} deserve comment.  An unsuspicious look
at Fig.\ \ref{fig_sigma05}(a) might suggest that the attractor
is quasiperiodic or chaotic for $0.52 \leq \kappa \leq 0.58,$ as
we see a complex set of data there. But this conclusion is premature,
as a highly-looped periodic attractor with many turning points
could also generate such a picture, thanks to the many local minima and
maxima in its corresponding time series.

Figure~\ref{fig_xy_proj} shows that, in fact, this is exactly what is
happening for many values of $\kappa$ in the  range
$0.52 \leq \kappa \leq 0.58$.
The $xy$ projections of the attractors in phase space are plotted using
\textsc{dstool} \cite{GMWW95}.  The orbit in Fig.\ \ref{fig_xy_proj}(a)
appears to be quasiperiodic, while those in
Figs.\ \ref{fig_xy_proj}(b)--(e) appear periodic.  These figures
also suggest that several kinds of
of bifurcations are taking place.  The trio
Figs.\ \ref{fig_xy_proj}(c)--(e) indicates a period-doubling sequence,
although we do not see the full period-doubling cascade to chaos.

Hence, to distinguish whether the long-term behavior is
periodic or not for a given set of parameters, a more careful
treatment, such as plotting the power spectrum, is needed.
We have computed the largest Lyapunov exponent as a function
of $\kappa,$ with other parameters fixed, and find that it is
always zero, up to  the accuracy of the numerical algorithms.
This result suggests that there is probably no
chaos in the intermediate coupling regime considered here.
Perhaps a more prudent way to put it is: chaos is not widespread for
intermediate coupling, if it occurs at all.

There are many other interesting attractors and bifurcations that
could be
discussed here, but we prefer to skip them so that we can focus our
attention on the self-similar picture near the unlocking threshold.


\section{A Singular Limit}
\label{section_A_Singular_Limit}

We now consider a simpler system that exhibits the same cascade found
in Fig.~\ref{fig_sigma05}.  Let  $\sigma=0$ in the reduced
system (\ref{receiver_sigma12}), so that it becomes
\begin{eqnarray}
   {d X_2 \over d t} &=& u X_2 + \kappa \cos \Phi,  \nonumber \\
   {d \Phi \over d t} &=& \delta - {\kappa \over X_2} \sin \Phi,
	     \nonumber \\
   {d u \over d t} &=& B - (1 + X_2^2).
\label{zero_sigma}
\end{eqnarray}

Figure~\ref{fig_singular} shows that this new system still
exhibits the striking pattern found earlier (although the
precise positions of the bifurcations are not the same, and there are
structures that come up in one case but not the other.)
This robustness suggests that the self-similar cascade is not due to
perturbative effects of $\sigma$ but is more generic.  To check that
the cascade is not just an artifact of the choice of variables
plotted, we have also tried plotting
local minima and maxima of $x, y, u$, and $I=x^2+y^2$ against the
parameter $\kappa.$  In all cases, we find similar patterns.
We have also checked that the patterns keep recurring even closer to
the unlocking threshold, specifically for $\kappa \in [0.998,1]$
and $\kappa \in [0.9998,1]$.  So we believe that the self-similarity
is genuine.

On the other hand, the limit $\sigma=0$ requires some caution on our
part.  The blown-up system (\ref{receiver_sigma12}) is obtained from
Eq.\ (\ref{receiver}) by the rescaling
$t = \sigma^{1/2} \tau, \sigma^{1/2} u = F_2 - 1, \Delta = \sigma^{1/2}
\delta, k = \sigma^{1/2} \kappa,$ which becomes {\it singular} as
$\sigma \rightarrow 0.$  Hence Eq.\ (\ref{zero_sigma}) is not directly
relevant physically, but it has the virtue that it leads to cleaner
numerics, and brings out the bifurcation mechanism more clearly.

One should also be aware that the singular limit displays some
nongeneric dynamics because of its higher degree of symmetry than the
original system.   For
instance, Eq.\ (\ref{zero_sigma}) is invariant under the
transformation
\begin{equation}
  {\mathcal T}: t \rightarrow -t, \ x \rightarrow -x,
	      \ y \rightarrow y, \ u \rightarrow -u,
\label{invariant_transformation}
\end{equation}
a ``reversibility'' symmetry that is not preserved by
Eq.~(\ref{receiver_sigma12}) for $\sigma \neq 0.$  A corollary of
this invariance is that if $\Omega$ is an attractor, then the set
$\tilde{\Omega} = \{ (x,y,u) | (-x,y,-u) \in \Omega \}$ is a repellor.
Results like this can be used for checking numerical convergence in
tracing bifurcation diagrams.

\subsection{Fixed-point analysis}
\label{subsection_Fixed_Points_for_Sigma0}

Since the transformation from Eq.\ (\ref{receiver}) to
Eq.\ (\ref{zero_sigma})
is singular, the fixed point analysis (both existence
and stability) in Sec.\ \ref{section_Locked_States}
is not directly applicable to Eq.\ (\ref{zero_sigma}).  The necessary
modification is straightforward, and the result is as
follows.  Fixed points exist if and only if
$1 \leq B \leq B_c \equiv 1 + {\kappa^2 / \delta^2}.$  They are given
by $p_{\pm} = (\pm x^*,y^*,\pm u^*),$ where
\begin{eqnarray*}
  x^* &=& {\delta\over\kappa} \sqrt{(B - 1)(B_c - B)}, \\
  y^* &=& {\delta\over\kappa} (B - 1), \\
  u^* &=& -\delta \sqrt{{B_c - B} \over {B - 1}}.
\end{eqnarray*}
Hence both fixed points recede to infinity as $B \rightarrow 1^+,$
and they coalesce at $(0, \kappa/\delta, 0)$ as
$B \rightarrow B_c^-.$

As for stability, the Jacobian of Eq.\ (\ref{zero_sigma}) is
\[ J = \left[ \begin{array}{lll}
		 u      & -\delta & x \\
		 \delta & u       & y \\
		 -2x    & -2y     & 0
	      \end{array}
	\right]. \]
At the fixed points, the characteristic equation simplifies to
\begin{equation}
  0 = f_{u}(\lambda) \equiv \lambda^3 - 2 u
  \lambda^2 + {\kappa^2 \over {B-1}} \lambda + 2 (B-1)(\lambda-u),
\label{CharEqn}
\end{equation}
where $u = u^*_{\pm}$ at the fixed point $p_{\pm}.$  Since
$f_{-u}(-\lambda) = - f_{u}(\lambda)$ and $u^*_+ = - u^*_-,$ we can
conclude that if
$\lambda$ is an eigenvalue of $p_+,$ then $-\lambda$ is an eigenvalue
of $p_-.$  Hence we cannot have a saddle-node
bifurcation at $B=B_c.$

The eigenvalue equation (\ref{CharEqn}), as it stands, is hard to
solve.  But some information can be obtained from the intermediate
value theorem.  Since
$f_{u}(\pm \infty) = \pm \infty,$ and
$f_{u^*_{\pm}}(0) = \pm 2 (B-1) \delta \sqrt{(B_c - B) / (B - 1)}$ is
positive (negative) at $p_+$ ($p_-$), the intermediate
value theorem implies that the fixed point $p_-$ always has a positive
eigenvalue and therefore cannot be stable.  Whether it is a saddle or
a repellor is not determined by this argument.

To investigate the type of bifurcation at $B=B_c,$ we will consider
Eq.\ (\ref{CharEqn}) with $B=B_c - \varepsilon,$ where
$\varepsilon > 0$ is a
small parameter, and $\kappa$ and $\delta$ are assumed to be $O(1)$
positive quantities.  Then we have a regular perturbation problem and
can find the eigenvalues to be
\begin{eqnarray}
   &&
   \lambda_{1,2} = \pm i \sqrt{\delta^2 + {2 \kappa^2 / \delta^2}}
		   - {{\delta^6 + \kappa^2 \delta^2} \over
		   {\kappa \delta^4 + 2 \kappa^3}} \sqrt{\varepsilon}
		   + O(\varepsilon), \nonumber \\
   &&
   \lambda_3 = {{-2 \kappa \delta^2} \over {\delta^4 + 2 \kappa^2}}
               \sqrt{\varepsilon} + O(\varepsilon)
\label{singular_eigenvalue:B}
\end{eqnarray}
at $p_+.$ Hence $p_+$ is an
attractor.  More specifically, it is an attracting ``spiral-node''.
The eigenvalues at $p_-$
can be obtained by taking the negatives of the eigenvalues of $p_+.$

These results provide some local information about the bifurcation at
$B=B_c$.  As $B \rightarrow B_c^-,$ i.e.,
$\varepsilon \rightarrow 0^+,$ the two fixed points approach each
other and collide at $B=B_c.$  Meanwhile, their attractiveness and
repulsiveness get weaker and weaker, since \emph{all}
the eigenvalues approach the imaginary
axis.  The fixed point at $B=B_c$ is linearly neutrally stable.

As Eq.\ (\ref{singular_eigenvalue:B}) indicates, locally the
bifurcation is of codimension two, with a simple zero eigenvalue and a
conjugate pair of pure imaginary eigenvalues.  As noted in
\cite{GH85}, such bifurcations can be complex because of
their global structures.  In Eq.\ (\ref{zero_sigma}), as in
Eq.\ (\ref{receiver_scaled}), the re-injection provides such a global
mechanism that leads to complicated dynamics.

Since the stable fixed point is annihilated in a collision with an
unstable object as $B \rightarrow {B_c}^-,$ we do not expect
stable objects for $B > B_c$ to be in small neighborhoods of
$(0,\delta (B_c - 1) / \kappa,0),$ the position of the fixed point
at $B=B_c.$  However, some form of intermittency might be expected,
and as we show in the next section,
this is indeed the case.  A trajectory on the periodic attractors
spends a long time in helical motion near the ghost of the fixed
point. The same is true when $\sigma>0,$ indicating that this helical
motion in the bottleneck is not caused by the degeneracy.

Before we move on to numerics, it should be noted that this degenerate
``attractor-repellor bifurcation'' is a consequence of setting
$\sigma$ to zero.  For $\sigma > 0,$ a generic saddle-node bifurcation
occurs at $B=B_c.$
Also, for $\sigma > 0,$ generically we do not have a genuine
codimension-two bifurcation.  But since both $H_1$ and $H_3$ in
(\ref{RouthHurwitz}) vanish at $B_c + O(\sigma),$ meaning that all
three eigenvalues are close to the imaginary axis when bifurcation
occurs, the nonsingular system (\ref{receiver_sigma12}) may pick up
some remnants of the bifurcation scenario of the singular system
(\ref{zero_sigma}), so that their orbit diagrams
(Figs.\ \ref{fig_sigma05} and \ref{fig_singular}) exhibit
similar features.  This suggests that whatever the bifurcation
scenario is, it is stable with respect to perturbations in parameters,
at least in an operational sense.

\subsection{Numerical studies of the self-similar bifurcation sequence}

In our numerical studies of the self-similar bifurcation sequence, we
concentrate on the periodic windows.  As before, we will vary
$\kappa$ and keep other parameters fixed.
Then for $\kappa = \kappa_c + \varepsilon,$ with
$0 \leq \varepsilon \ll 1$ and $\kappa_c \equiv \delta \sqrt{B-1},$
the fixed points are at $p_{\pm} = (\pm x^*,y^*,\pm u^*),$ where
\begin{eqnarray*}
  x^* &=& {1\over\delta} \sqrt{2 \varepsilon \kappa_c} (1 -
	  {3 \varepsilon \over 4 \kappa_c}) + O(\varepsilon^{5/2}), \\
  y^* &=& {\kappa_c\over\delta} (1 - {\varepsilon \over \kappa_c})
	  + O(\varepsilon^2), \\
  u^* &=& - {\delta\over\kappa_c} \sqrt{2 \varepsilon \kappa_c}
	  (1 + {\varepsilon \over 4 \kappa_c}) + O(\varepsilon^{5/2}).
\end{eqnarray*}
The eigenvalues at $p_+$ are
\begin{eqnarray}
   &&
   \lambda_{1,2} = \pm i \sqrt{\delta^2 + {2 {\kappa_c}^2 / \delta^2}}
		   - \delta \sqrt{2\over\kappa_c}
		   {{\delta^4 + {\kappa_c}^2} \over
		   {\delta^4 + 2 {\kappa_c}^2}} \sqrt{\varepsilon}
		   + O(\varepsilon), \nonumber \\
   &&
   \lambda_3 = {{- \delta (2 \kappa_c)^{3/2} } \over
	       {\delta^4 + 2 {\kappa_c}^2}}
               \sqrt{\varepsilon} + O(\varepsilon).
\label{singular_eigenvalue:kappa}
\end{eqnarray}

\subsubsection{Helical structure of the periodic attractors}
\label{subsubsection_SPA}

Using \textsc{auto}, we have examined the phase-space geometry of the
attractors in individual periodic windows.  For definiteness,
consider the $yu$ projections of the attractors.
Figure~\ref{fig_n7window} shows a typical collection of snapshots
as we sweep $\kappa$ across a window.  What happens in this window
also happens in other windows, with some modifications in details.
Specifically, we see $n$ little helical loops per round trip
at the left end of the $n$th window.  (In this example, $n=7$.)
Most of these loops are located near $(0,1,0),$
where the stable fixed point
is annihilated at $\kappa=1.$

Intuitively, the trajectory is slowly funneled through the bottleneck
caused by the ghost of the former stable fixed point.  During this slow
passage, the trajectory also spirals around at an $O(1)$ frequency, 
given by the imaginary part of the complex eigenvalues in
Eq.\ (\ref{singular_eigenvalue:kappa}).  The combination of slow 
passage and $O(1)$ spiralling gives rise to the helical appearance 
of the trajectories.

As we increase $\kappa$ in a periodic window,
$(n-2)$ of these loops stay at roughly the same distance from the
$u$ axis, while the remaining two loops are stretched in
the re-injection part of the flow.  Moreover,
the left side of the projection moves, with a stronger shear at
the bottom, towards the
$u$ axis, eventually crossing the loops.  Meanwhile,
there is a clockwise rotation to bring about the
structure we see as we reach the right-hand end of the window.
The same trend occurs in all the periodic windows we have sampled,
with $3
\leq n \leq 12$ (we do not see periodic windows corresponding
to $n=1,2$).
If this ``$n$ loops in the $n$th window'' scenario is correct,
and if there really exist infinitely many such windows accumulating
at the limiting value $\kappa=1,$ we will have an attractor
with infinitely many tight loops at $\kappa=1^-$, just before its
annihilation in an
attractor-repellor bifurcation.  This will then be reminiscent of
the chaotic orbit born from the infinite sequence of period doubling in
the logistic map.

In a similar problem concerning a $\mathrm{CO_2}$ laser, Zimmermann
{\it et al.}\ \cite{ZNS97} worked backwards and constructed a model
by assembling a local spiral motion and a global re-injection.
Numerically, for a fixed set of parameters, the
flow is similar to what we see here.  However, how this mathematical
model is related to the original laser system is unclear.

\subsubsection{How periodicity ends}

In each periodic window, the unstable branch always has its nontrivial
Floquet multipliers outside the unit circle,
indicating a repellor
instead of a saddle.  When this branch collides with the stable branch,
an attractor-repellor bifurcation of cycles occurs, bringing an end to
the periodic window.  In virtue of the invariance of
Eq.\ (\ref{zero_sigma}) under the
transformation $\mathcal T$ in (\ref{invariant_transformation}),
it is no wonder
that at the bifurcations, when the attractor and the repellor coalesce,
this neutrally stable object is invariant under
${\mathcal T'}: x \rightarrow -x, y \rightarrow y, u \rightarrow -u.$
This explains why the projections onto the $yu$ plane are always
symmetrical with respect to $u \rightarrow -u$
[as seen in Figs.\ \ref{fig_n7window}(a) and (d)],
while the projections
onto the $xy$ plane are always
symmetrical with respect to $x \rightarrow -x$ at the ends of the
windows (not shown).

As mentioned earlier, the attractor-repellor bifurcations are
a consequence of the reversibility symmetry of the $\sigma=0$ system.
Generically, the periodic attractors are annihilated in saddle-node
bifurcations and there is no $\mathcal T$-symmetry, a strong
indication that the
mechanism underlying the cascade of bifurcations has nothing to do with
this symmetry.

Recalling the results in
Sec.\ \ref{subsection_Fixed_Points_for_Sigma0},
we notice that while an attractor-repellor bifurcation of fixed points
occurs as the locking threshold is approached from above ($\kappa
\rightarrow 1^+)$, there is a simultaneous attractor-repellor
bifurcation of cycles (presumably with infinitely many loops) as
the threshold is approached from below ($\kappa \rightarrow 1^-).$
This double-sided aspect
of the unlocking bifurcation also seems to be preserved for $\sigma>0,$
suggesting that it too has nothing to do with symmetry.

\subsubsection{Trends in the periodic windows}

The periodic windows in Fig.\ \ref{fig_singular}
have been verified by  \textsc{auto} to be isolas,
i.e., they terminate at both ends as the periodic attractors are
annihilated in collisions with unstable objects (probably unstable 
periodic orbits).
However, we do not fully understand the mechanism that causes such
annihilations, although it seems to be some form of resonance of the
spiralling motion and the global re-injection.  Indeed, we have found
that as we increase $\kappa$ in a periodic window, periodicity ends
precisely when a trajectory executes an integer number of spiral loops
upon one re-injection, i.e., if
\begin{equation}
  {2 \pi \over \omega_I} = n {2 \pi \over \omega_S},
\label{resonance_condition}
\end{equation}
where $n$ is a positive integer, and $\omega_I$ and $\omega_S$ are the
frequencies of the re-injection and the spiralling motion,
respectively.
To see this, we assume that the time needed for re-injection is
dominated by the slow passage through the bottleneck.  Then, using
Eq.\ (\ref{singular_eigenvalue:kappa}), we may rewrite
Eq.\ (\ref{resonance_condition}) as
\[
   {\delta (2 \kappa_c)^{3/2} \over \delta^4 + 2 {\kappa_c}^2}
   \sqrt{\varepsilon_n} =
   {1 \over n} \sqrt{\delta^2 + {2 {\kappa_c}^2 / \delta^2}}
\]
to lowest order, where $\varepsilon_n = 1 - \kappa_{\rm R,n},$ and
$\kappa_{\rm R,n}$ is where the $n$th periodic window ends at the
right.  Solving for $\varepsilon_n,$ we get
\begin{equation}
  \varepsilon_n = {(\delta^4 + 2 {\kappa_c}^2)^3 \over \delta^4
		  (2 \kappa_c)^3} {1 \over n^2}.
\label{bifn_position}
\end{equation}
Moreover, at these parameter values, corresponding to the right ends of
the periodic windows, the periods of the attractors are expected to be
\begin{equation}
  T_{\rm R,n} \approx {2 \pi \over \omega_I}
  = 2 \pi {\delta^4 + 2 {\kappa_c}^2 \over \delta (2 \kappa_c)^{3/2}}
    {1 \over \sqrt{\varepsilon_n}}
  = {2 \pi \delta \over \sqrt{\delta^4 + 2 {\kappa_c}^2}} n.
\label{period_right}
\end{equation}

Figures \ref{fig_trends}(a) and (b) compare these predicted scaling 
laws against numerics.  For $n \geq 5,$ Eqs.\ (\ref{bifn_position}) 
and (\ref{period_right}) agree with numerics to within $2\%.$

We do not know what mechanism kills the periodic window as we decrease
$\kappa.$ But drawing analogy to the mechanism at the other end, we
expect that the positions of the left ends of the periodic windows
scale as
\[ 1 - \kappa_{\rm L,n} = O(1/n^2), \]
so that the widths of the periodic windows scale as
\begin{equation}
  w_n = \kappa_{\rm R,n} - \kappa_{\rm L,n} = O(1/n^3),
\label{width}
\end{equation}
and the periods of the attractors at the left ends of the periodic
windows scale as
\begin{equation}
  T_{\rm L,n} = O(n).
\label{period_left}
\end{equation}
These scaling behaviors are verified in
Figs.\ \ref{fig_trends}(c)--(d).

Meanwhile, the sizes of the attractors, as
measured by the $L^2$-norms, remain $O(1)$
as $\kappa \rightarrow 1^-.$
This is expected as the $L^2$-norm is dominated by the global
re-injection, which persists in all periodic windows.


\section{Open Questions}
\label{section_Open_Problems}

From a theoretical perspective, the most interesting open question
concerns the mathematical mechanism underlying the self-similar cascade
of bifurcations observed in both Eq.\ (\ref{receiver_sigma12})
and the simpler system (\ref{zero_sigma}).  Whatever the mechanism is, 
the heuristic arguments and numerical evidence presented in 
Sec.\ \ref{section_A_Singular_Limit} suggest that it must combine the 
features of a saddle-node infinite-period bifurcation and
a supercritical Hopf bifurcation.  As such, it may well arise in
other scientific settings.
Maybe it can even be detected experimentally.

Although self-similarity itself is common in dynamical systems,
and has been explained by renormalization group arguments in
such contexts as period doubling, intermittency, and quasiperiodic
breakdown \cite{GH85}, it seems the cascade we see here
falls into none of these
categories.  Rather, it is characterized by an infinite series of
saddle-node bifurcations of cycles, accumulating at a finite parameter
value corresponding to the locking threshold.

Aside from these bifurcation issues, the dynamics of
Eq.\ (\ref{zero_sigma}) is also interesting for a {\it fixed}
set of parameters.
Zimmermann {\it et al.}\ \cite{ZNS97} suggested that the periodic
orbits with many small loops can be understood as the effects
of a flow with helical local dynamics together with a global
re-injection.  The challenge now is to find a way to reduce the laser
equations to a form where this conjectured phase space geometry
becomes transparent.

There are many other interesting avenues for future research.
The dynamics of the unidirectionally forced system could be explored
over a much broader range of parameter values, with $k$ and $\Delta$
not necessarily of the same order.  We have also neglected the
effects of noise, a topic of great importance in technological
applications of injection locking
\cite{Barillet96,BTRHWM95,FW94,FGB95}.  Another promising direction
would be to study arrays of coupled lasers
\cite{BKWK95,KBKW98,SFW93,LE92,LE94,BZS97,HGEK97}
driven by external injection, particularly in regimes where amplitude
effects are important
and the phase model approximation is not valid.  Finally, we have
restricted attention to drive signals of constant intensity and
frequency,
but for applications to optical communications \cite{CR94,VR98},
one needs to study how lasers respond to modulated drive
signals, especially those carrying messages within them.

\acknowledgements
We thank:  John Guckenheimer, for discussions on period doubling
and on the
self-similar bifurcation sequence; Jim Keener, for pointing out that
essential singularities can give rise to self-similar bifurcation
sequences;  Rajarshi Roy and Scott Thornburg, for discussions on the
physical interpretation of our results and for references to the laser
literature; and Paul Steen, for commenting on an early draft of this
article and for providing technical advice regarding \textsc{auto}.
We also thank Henry Abarbanel, Rajarshi Roy, and the other members of
the UCSD/Georgia Tech/Cornell collaboration on synchronization and
communication in nonlinear optical systems.
Research supported in part by the National Science Foundation.

%


\begin{thebibliography}{10}
\bibitem{rectangular_cylindrical}
Alternatively, we may use rectangular coordinates, with 
  $x = X_2 \cos \Phi, y = X_2 \sin \Phi$ to avoid the notational 
  difficulty regarding the case $X_2 = 0.$ 
  The results are of course the same.
\bibitem{no_collision}
At $B = B_c = 1 + \kappa^2/\delta^2,$ there are three, not two, 
  fixed points.
  They are at $1 + \kappa^2 / (2 \delta^2) \pm \sqrt{\kappa^4 / 
  (4 \delta^4) -
  \sigma^{2a} \delta^2 - \sigma^{2a} \kappa^2}$ and $1.$ These are 
  equivalent
  to $B_c - \sigma^{2a} \kappa^2 B_c / (B_c - 1)^2 + O(\sigma^{4a}),$ 
  $1 + \sigma^{2a} \kappa^2 B_c / (B_c - 1)^2 + O(\sigma^{4a}),$ 
  and $1$ (exact), in
  agreement with Eqs.\ (\ref{fixed_point_1}) and 
  (\ref{fixed_point_23}). There
  is no collision of fixed points at $B=B_c$.
\bibitem{Routh_Hurwitz_bifn}
If we start from a set of parameters such that (\ref{RouthHurwitz}) 
  holds and
  vary parameters continuously, we cannot have $H_2 = 0$ or 
  $H_{2'} = 0$
  without having one of the other two inequalities becoming an 
  equality first.
  Hence we can concentrate on $H_1$ and $H_3$ when we look for 
  bifurcation
  points.
\end{thebibliography}
%

%
%

\section*{List of Figure Captions}

\begin{figure}[H]
  \caption{
  The dependence of the behaviors of the phase difference $\Phi$ on
    the pump to loss ratio $B,$ for
    $k \sim \Delta = O(\sigma^a), a > 1/2.$
    $\Delta, k$ are as defined in Eq.\ (\ref{Delta_and_k}).
    $E = X_2 e^{i \Phi}$ is the complex electric field.
    The stable (unstable) fixed point is represented by a
    ``$\bullet$'' (``$\circ$'').
    $\Phi$ is constant if
    $B < B_c \equiv 1 + k^2/\Delta^2$ and increases strictly if
    $B > B_c,$ with a saddle-node infinite-period bifurcation at 
    $B = B_c.$
    The bifurcation value for $B$ has $O(\sigma^{2a})$ errors
    [Eq.\ (\ref{RH1})] and the value of $X_2$ has $O(\sigma^a)$ errors
    [Eq.\ (\ref{phase_model_soln})].  Contrast this with
    Fig.\ \ref{fig_hopf}. }
\label{fig_saddle_node}
\end{figure}

\begin{figure}[H]
  \caption{
  The dependence of the behaviors of the phase difference $\Phi$ on
    $B,$ for $k \sim \Delta = O(\sigma^a), a < 1/2.$  Parameters,
    variables, and symbols are as defined in
    Fig.\ \ref{fig_saddle_node}.
    $\Phi$ is constant if
    $B < B_c,$ oscillates between ${\pi / 2}
    \pm \tan^{-1} ({\Delta A^{(0)} / k})$ if $B_c < B < 1 + 2 k^2 /
    \Delta^2,$ and increases strictly if
    $B > 1 + 2 k^2 / \Delta^2.$  $A^{(0)}$ is as defined in
    (\ref{Hopf_amplitude}).
    A supercritical Hopf bifurcation occurs at $B = B_c.$
    The bifurcation value for $B$ has $O(\sigma^{2-2a})$ errors
    [Eq.\ (\ref{RH3})] and the value of $X_2$ has $O(\sigma^{1-2a})$
    errors [Eq.\ (\ref{hopf_rescaled_result})].  Contrast this with
    Fig.\ \ref{fig_saddle_node}, where phase trapping is
    impossible. }
\label{fig_hopf}
\end{figure}

\begin{figure}[H]
  \caption{
  Orbit diagram, plotting the local minima and maxima of the intensity
    $I = X_2^2$ vs $k,$ of Eq.\ (\ref{receiver}),
    with $\sigma=0.010, \Delta=2.0,
    k \in [0,2.2], B=2.0.$ Dots represent numerical results and curves
    theoretical expectations [Eq.\ (\ref{twotimingresults})].
    In this and subsequent orbit diagrams
    (Figs.\ \ref{fig_sigma05} and \ref{fig_singular}), the
    numerical results are obtained by following the equation of motion
    with the subroutine \textsc{stiff} in
    [W. H. Press, S. A. Teukolsky, W. T. Vetterling and
    B. P. Flannery, \textit{Numerical Recipes in Fortran:
    The Art of Scientific Computing} (Cambridge University Press,
    New York, 1992), 2nd ed.], which
    implements a fourth order Rosenbrock method, and then estimating
    the extremal values of the intensity with a second order
    interpolation.  The results are corroborated by another
    subroutine, \textsc{stifbs} [\textit{ibid.}], which implements a
    semi-implicit extrapolation algorithm analogous to the
    Bulirsch-Stoer method.
    }
\label{fig_order1}
\end{figure}

\begin{figure}[H]
  \caption{
  Numerically computed orbit diagram of Eq.\ (\ref{receiver_sigma12}),
    with  $\sigma=0.0025, \delta = 1.0, B=2.0.$
    (a) $\kappa \in [0.40,1.1];$
    (b) a blow-up in the range $\kappa \in [0.98,1.0].$
    Contrast the complicated patterns with the simple structure in
    Fig.\ \ref{fig_order1}.  Note in particular the self-similar
    bifurcation sequence that piles up at
    $\kappa = 0.9986 = \delta \sqrt{B-1} + O(\sigma).$
    }
\label{fig_sigma05}
\end{figure}

\begin{figure}[H]
  \caption{
  The $xy$ projections of the attractors of
    Eq.\ (\ref{receiver_sigma12}),
    with $\sigma=0.0025, \delta = 1.0, B=2.0.$
    (a) $\kappa=0.78,$ a quasiperiodic attractor.  We have followed
        the flow only for a time span of $1000$ units in order to
	reveal the structure.
    (b) $\kappa=0.77,$ a periodic attractor with many local minima
        and maxima.  This attractor remains a closed curve without
	blurring in a time span of $2000$ units, strongly hinting
	its periodicity.
    (c) $\kappa=0.70,$ a period-1 orbit.
    (d) $\kappa=0.68,$ a period-2 orbit.
    (e) $\kappa=0.67,$ a period-4 orbit.
    }
\label{fig_xy_proj}
\end{figure}

\begin{figure}[H]
  \caption{
  Numerically computed orbit diagram of Eq.\ (\ref{zero_sigma}),
    with $\delta=1.0,B=2.0.$
    (a) $\kappa \in [0,1.1];$
    (b) $\kappa \in [0.98,1.0].$
    Compare with Fig.\ \ref{fig_sigma05}.
  }
\label{fig_singular}
\end{figure}

\begin{figure}[H]
  \caption{The $yu$ projections of the attractors of
  Eq.\ (\ref{zero_sigma}) as we sweep
  across the $n=7$
  periodic window by increasing $\kappa,$ with $\delta=1.0, B=2.0.$
  (a) $\kappa=0.91867;$
  (b) $\kappa=0.92500;$
  (c) $\kappa=0.92980;$
  (d) $\kappa=0.93200.$
  The snapshots are not taken at evenly spaced values of
  $\kappa.$  Rather, representatives are chosen to show
  the deformations more clearly.
  }
\label{fig_n7window}
\end{figure}

\begin{figure}[H]
  \caption{
  Log-log plots showing the
  trends among the periodic windows of Eq.\ (\ref{zero_sigma}) with
    $\delta=1.0, B=2.0,$ in the
  (a) positions of the right ends of the periodic windows, with the
      theoretical prediction $\ln (1 - \kappa_{\rm R,n}) =
      \ln (27/8) - 2 \ln n;$
  (b) periods of the attractors at the right ends of the
      periodic windows, with the theoretical prediction
      $\ln T_{\rm R,n} = \ln (2 \pi / \sqrt{3}) + \ln n;$
  (c) widths, with the fit $ \ln w_n = 1.5 - 3 \ln n;$ and
  (d) periods of the attractors at the left ends of the
      periodic windows, with the fit
      $\ln T_{\rm L,n} = 1.2 + \ln n.$ 
  Dots represent raw data with $3 \leq n \leq 12,$
  solid lines represent theoretical predictions
  [Eqs.\ (\ref{bifn_position}) and (\ref{period_right})]
  with no fitting parameter, and dashed
  lines represent semi-theoretical predictions, where the slopes are
  chosen according to the theoretical scaling
  [Eqs.\ (\ref{width}) and (\ref{period_left})],
  and the intercepts are chosen to fit the data.
  }
\label{fig_trends}
\end{figure}

%
%

\section*{Figures}

\begin{figure}[htp]
\centerline{\epsffile{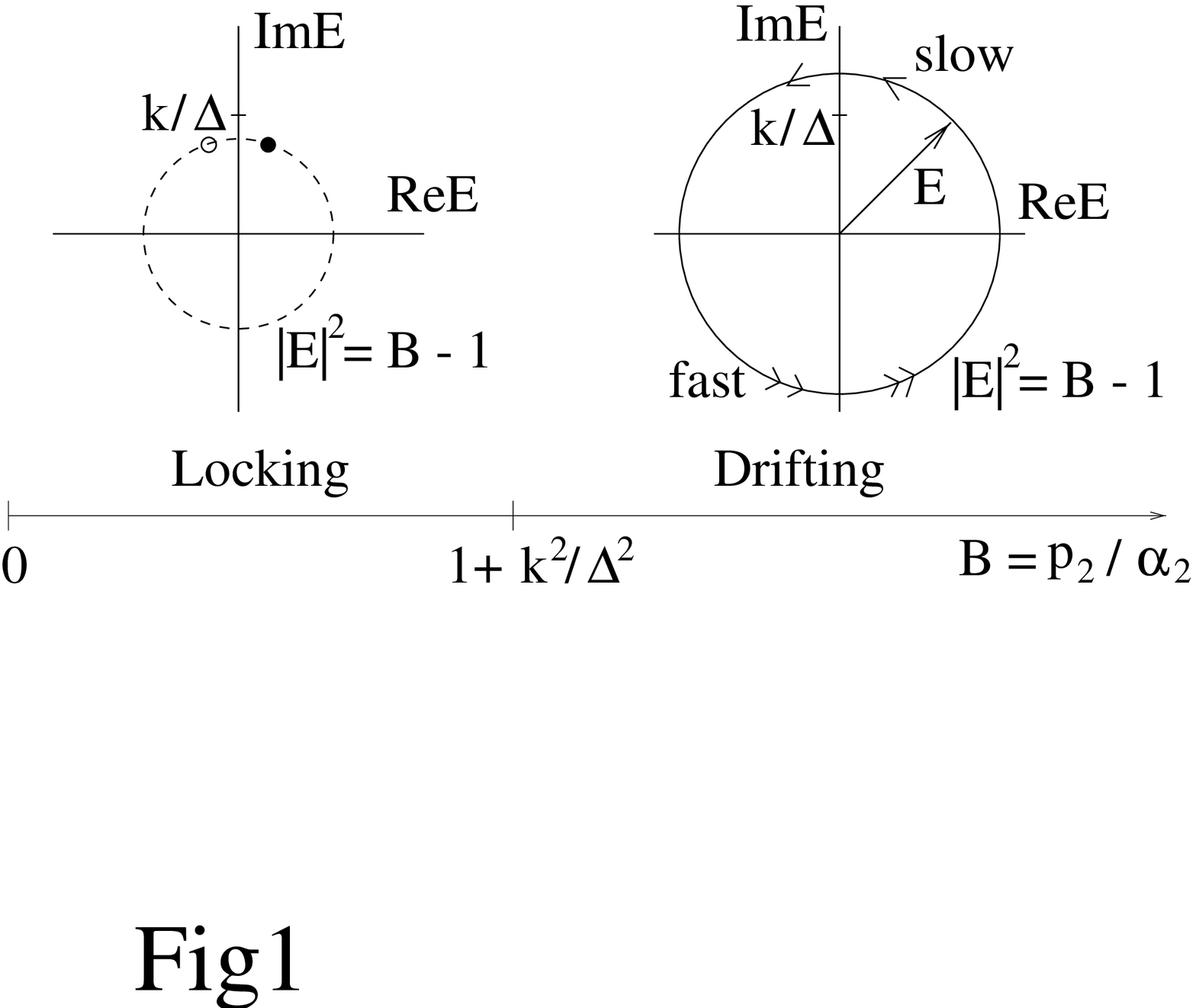}}
\end{figure}

\begin{figure}[htp]
\centerline{\epsffile{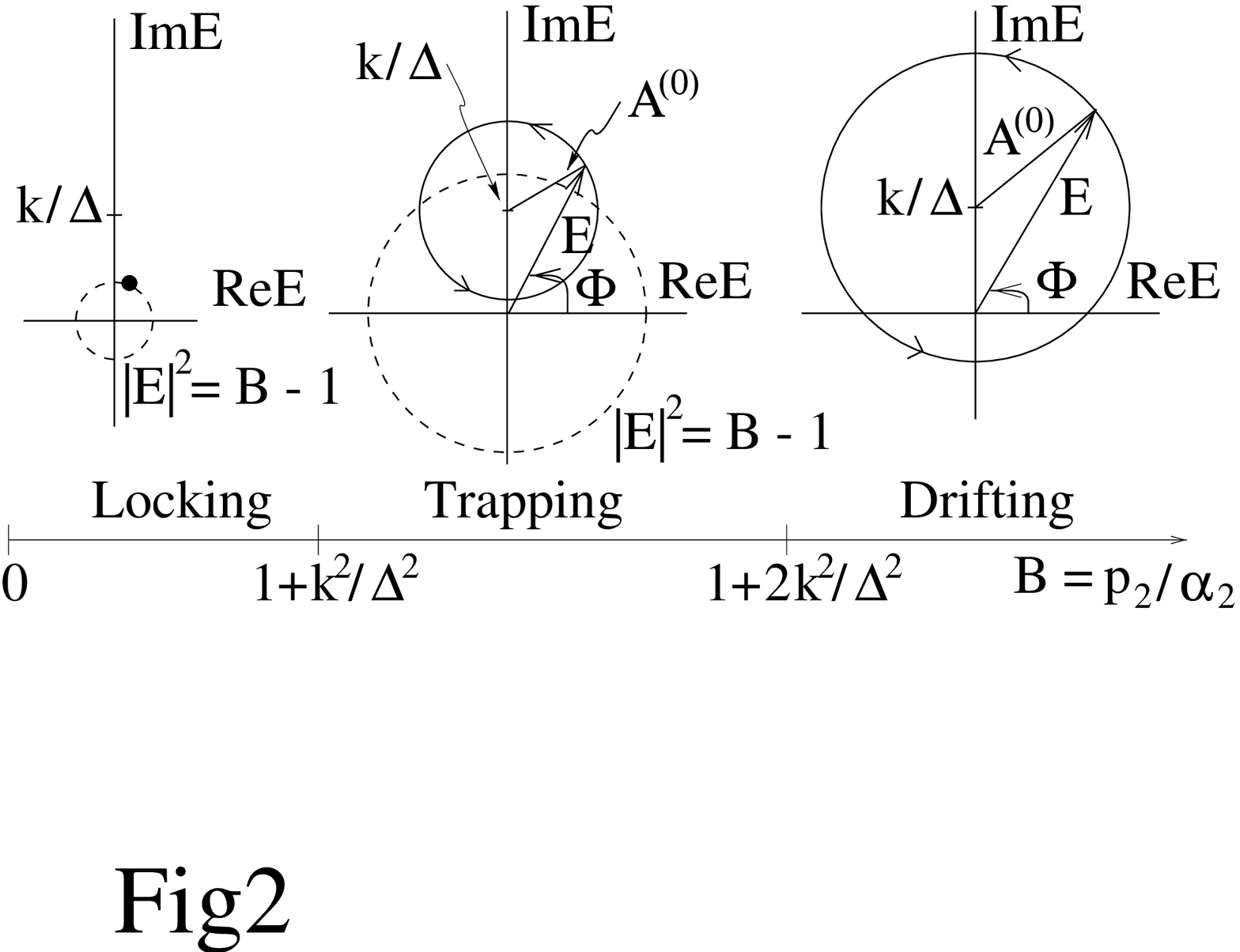}}
\end{figure}

\begin{figure}[htp]
\centerline{\epsffile{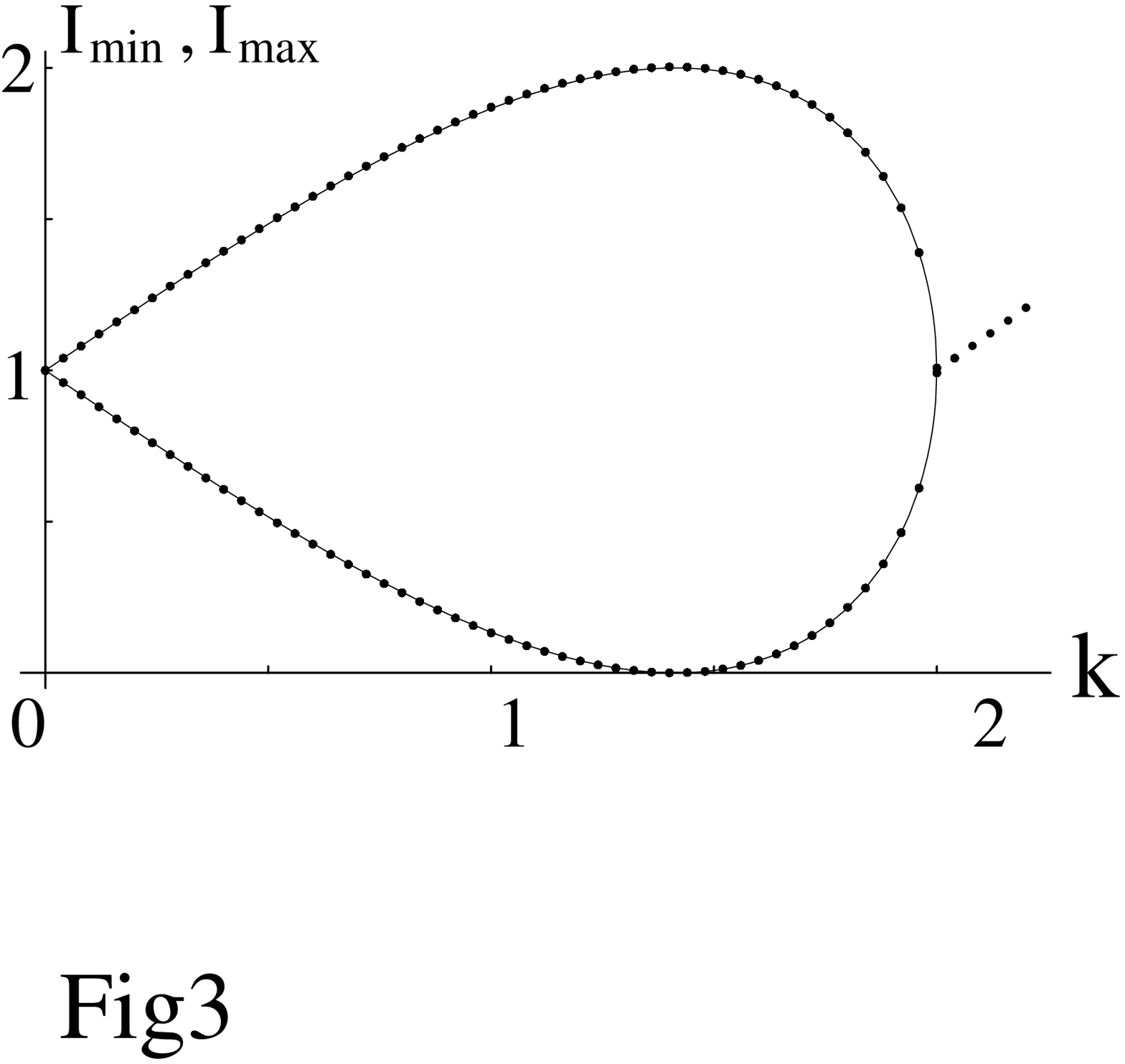}}
\end{figure}

\begin{figure}[htp]
\centerline{\epsffile{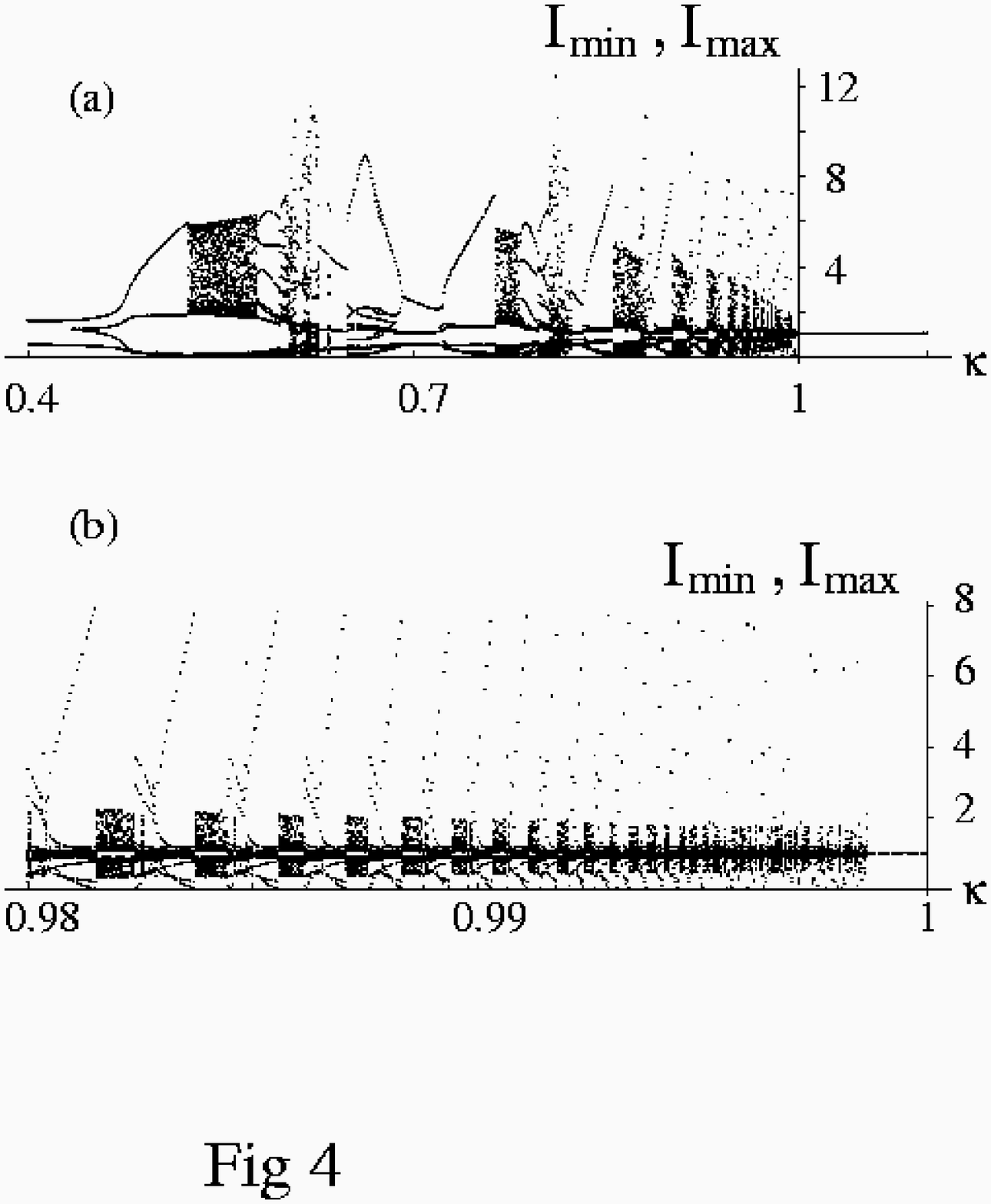}}
\end{figure}

\begin{figure}[htp]
\centerline{\epsffile{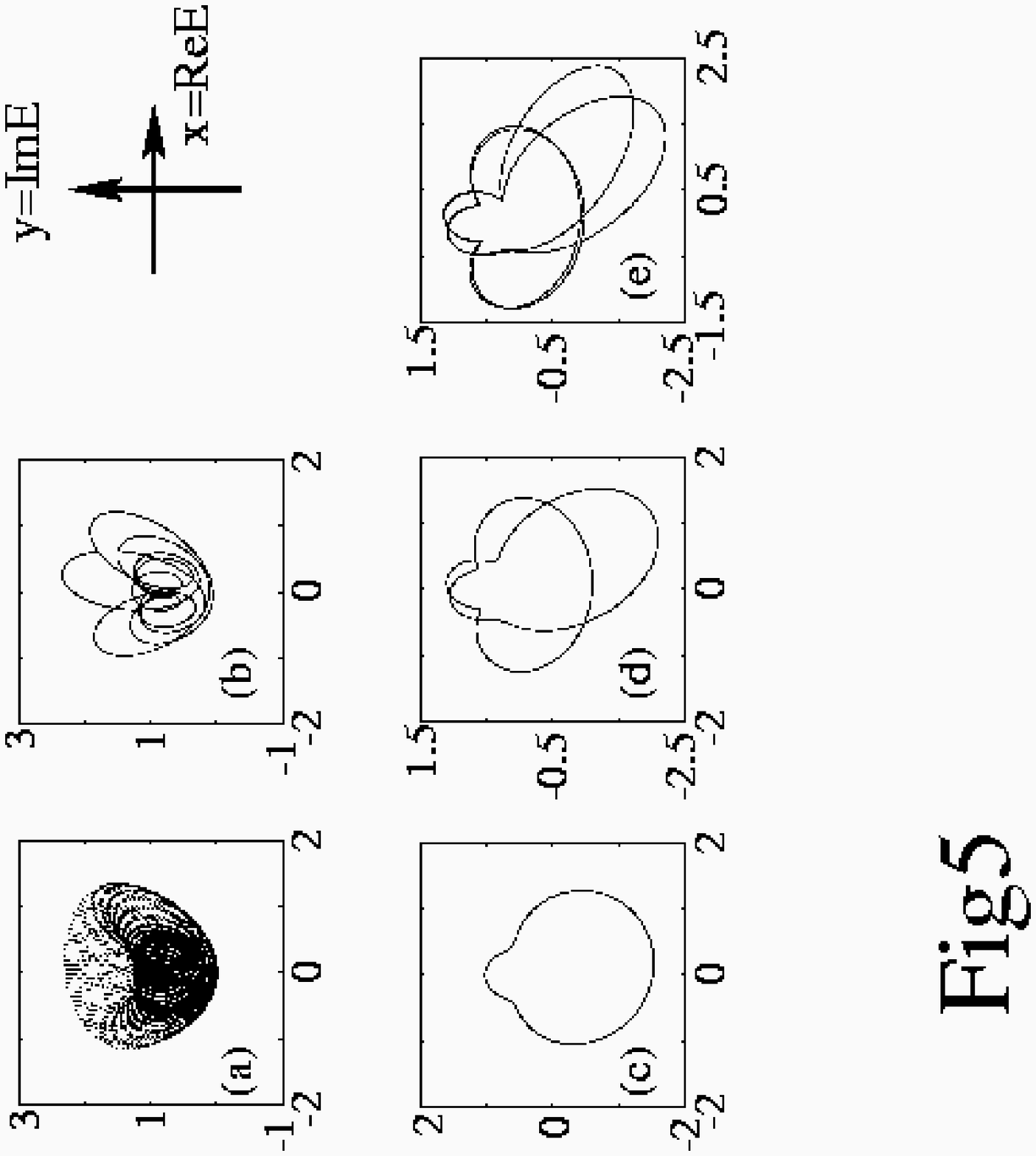}}
\end{figure}

\begin{figure}[htp]
\centerline{\epsffile{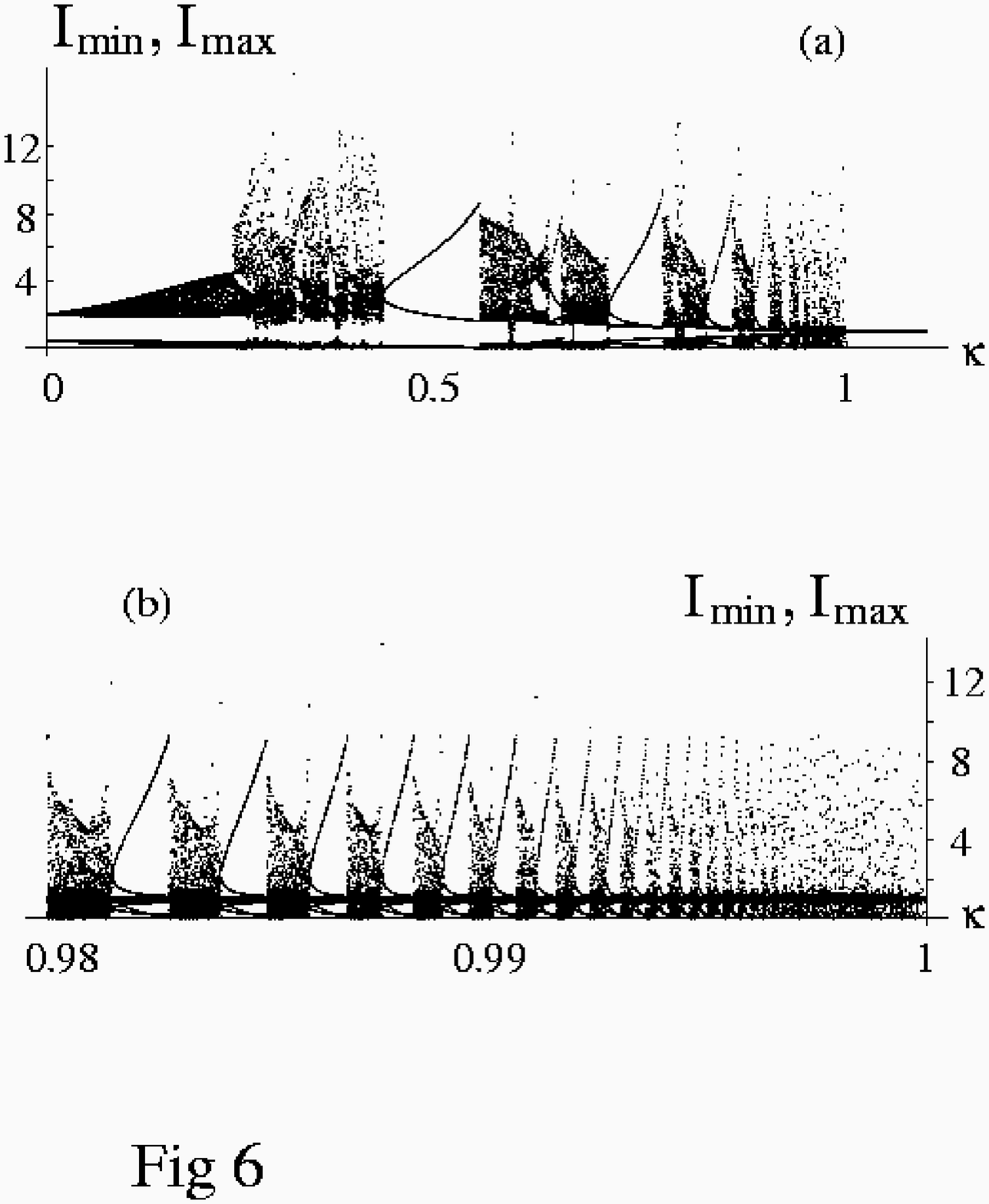}}
\end{figure}

\begin{figure}[htp]
\centerline{\epsffile{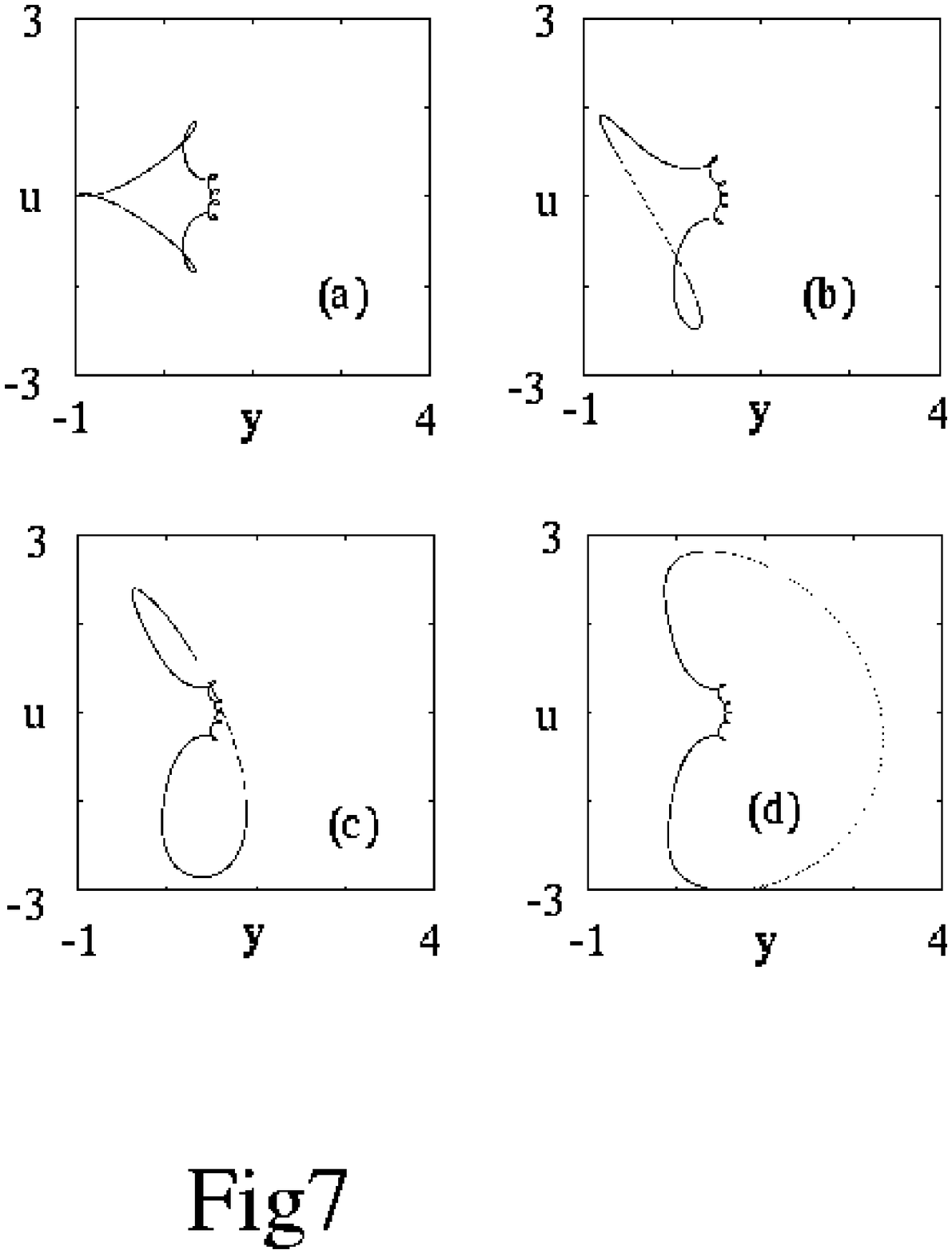}}
\end{figure}

\begin{figure}[htp]
\centerline{\epsffile{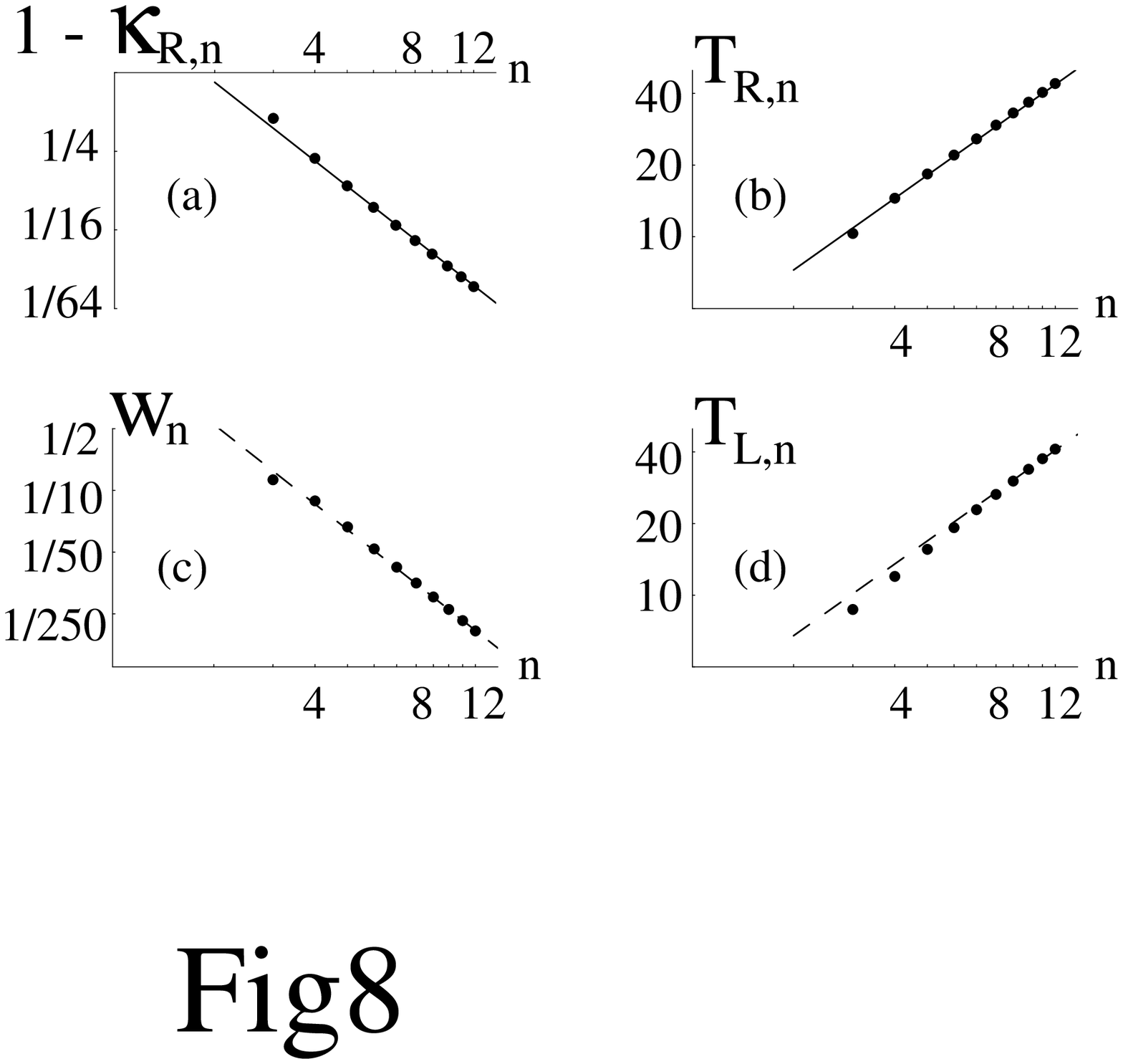}}
\end{figure}

%
%

\end{document}